\documentclass[aps,prb,reprint,superscriptaddress]{revtex4-1}
\usepackage[pdftex]{graphicx}
\usepackage{color}
\usepackage{changes}
\usepackage{amsmath}

\begin{document}


\title{Magnetocrystalline Anisotropy of Fe-based $L1_0$ Alloys: 
Validity of Approximate Methods to Treat the Spin-Orbit Interaction}



\newcommand{\QUIM}[0]{{
Departamento de F\'{\i}sica de Materiales, Facultad de Qu\'{\i}mica, 
Universidad del Pa\'{\i}s Vasco UPV/EHU,
Apartado 1072, 20080 Donostia-San Sebasti\'an, Spain}}

\newcommand{\DIPC}[0]{{
Donostia International Physics Center, 
Paseo Manuel de Lardiz\'abal 4, 20018 Donostia-San Sebasti\'an, Spain}}

\newcommand{\CFM}[0]{{
Centro de F\'{\i}sica de Materiales CFM/MPC (CSIC-UPV/EHU), 
Paseo Manuel de Lardiz\'abal 5, 20018 Donostia-San Sebasti\'an, Spain}}

\newcommand{\ICMM}[0]{{
Instituto de Ciencia de Materiales de Madrid, CSIC,
Cantoblanco, 28049 Madrid, Spain}}

\affiliation{\QUIM}
\affiliation{\DIPC}
\affiliation{\DIPC}
\affiliation{\ICMM}

\author{M. Blanco-Rey}
\email[]{maria.blanco@ehu.es}
\affiliation{\QUIM}
\affiliation{\DIPC}

\author{J.I. Cerd\'a}
\affiliation{\ICMM}

\author{A. Arnau}
\affiliation{\CFM}
\affiliation{\QUIM}
\affiliation{\DIPC}


\date{\today}

\begin{abstract}
First-principles calculations are used to gauge different levels of approximation to calculate
the magnetocrystalline anisotropy energies (MAE) of five $L1_0$ FeMe alloys (Me=Co, Cu, Pd, Pt, Au).
We find that a second-order perturbation (2PT) treatment of the spin-orbit interaction (SOI)
breaks down for the alloys containing heavier ions, while it provides a very accurate
description of the MAE behaviour of FeCo, FeCu, and FePd.
Moreover, the robustness of  the 2PT approximation is such that in these
cases it accounts for the MAE of highly-non-neutral alloys and, thus,
it can be used to predict their performance when dopants are
present or when they are subject to applied gate bias, which
are typical conditions in working magnetoelectric devices.
We also observe that switching of the easy axis direction can be induced in some of these alloys
by addition or removal of, at least, one electron per cell.
In all cases, the details of the bandstructure are responsible for the finally observed MAE value
and, therefore, suggest a limited predicting power of models based on
the expected orbital moment values and bandwidths. 
Finally, we have confirmed the importance of various calculation parameters
to obtain converged MAE values, in particular,
those related to the accuracy of the Fermi level determination.
\end{abstract}


\maketitle

\section{Introduction}

In a model system of interacting magnetic moments various
contributions can be identified that lead to the observation of the
magnetic anisotropy, i.e. the existence of a preferential
magnetization direction in the system. These terms are the
classical dipole-dipole interaction, resulting in the so-called
shape anisotropy, and quantum-mechanical ones, as 
anisotropic exchange \cite{bib:moriya60}
and the magnetocrystalline
anisotropy (MCA). The latter, of relativistic character, has its
origin in the electronic spin-orbit interaction (SOI).

The historical development of non-volatile memories based in
the property of magnetorresistivity has been closely related to
the ability of balancing out those competing contributions.
The key achievement was the perpendicular magnetic anisotropy
in thin film heterostructures, since the out-of-plane MCA at the
interface tends to dominate in-plane shape anisotropy.
The efficient spin orientation control in the ferromagnetic
(FM) electrodes of magnetic tunneling junctions is still a technological
challenge. The use of external magnetic fields for this purpose is evolving 
{towards voltage control of spintronic devices \cite{bib:wang12,bib:pantel12}
and spin transfer torques induced by spin-polarized currents \cite{bib:dieny17}.
These advances motivate efficient, robust and accurate 
modelling of the physics of the MCA for different materials/interfaces under
external stimuli, such as external fields or strain.

The interplay of MCA and dimensionality is considered as a promising
route for the development of spintronics. 
Indeed, the MCA plays a central role in the magnetic
properties of two-dimensional materials, since it can prevent
thermal fluctuations from destroying long-range magnetic
order \cite{bib:mermin66}.
In solids with cubic symmetry, the MCA is an effect of fourth order in
the SOI strength, but a tetragonal distortion can enable a second-order MCA.
This idea is behind the materials used in (or proposed for)
some of the aforementioned devices. For example, multiferroics
allow for strain-mediated magneto-electric coupling \cite{bib:jia12,bib:duan08b} and,
more recently, tetragonal Heusler alloys have attracted attention for combining
high MCA and half-metallicity \cite{bib:faleev17,bib:herper18}.
We put the focus of this work on transition-metal alloys. They are
versatile, since their structure and magnetic properties can be
tailored by varying the stoichiometry, as it is the case of the
strongly magnetostrictive ``galfenol'' (Fe$_{1-x}$Ga$_x$) \cite{bib:zhang10,bib:wang13},
Fe$_{1-x}$Co$_x$ \cite{bib:burkert04,bib:andersson06,bib:turek12}, and Cu-Ni 
films \cite{bib:quintana17}.
Several theoretical studies have shown that a bias voltage could significantly affect the
MCA \cite{bib:duan08,bib:nakamura09} and, in fact, an electric-field-induced MCA switching 
has been realized in Fe$_{30}$Co$_{70}$ alloy films \cite{bib:gamble09}.


Density Functional Theory (DFT) calculations that include SOI provide
information on the magnetocrystalline anisotropy energy (MAE).
Nonetheless, its small magnitude, often in the sub-meV range, 
imposes stringent convergence to the DFT calculations at the expense
of high computational demands.
In practice, fully relativistic calculations that include SOI
are carried out by first computing self-consistently the Hamiltonian of
the system including the scalar relativistic term
and next adding the spin-orbit contribution. 
The latter may be treated 
self-consistently (SCF) or non-self-consistently~\cite{bib:daalderop90,bib:li90}
(NSCF) within the so-called force theorem \cite{bib:weinert85} to obtain the MAE.
NSCF is often considered to produce a good estimate.
Other electronic structure calculations use the
second-order perturbation theory (2PT) to calculate MAEs and
orbital moments. These alternative approaches rely on the
knowledge of the spin-orbit effects on atomic orbitals
\cite{bib:bruno89,bib:cinal94,bib:vanderlaan98}
and have been formally formulated in the literature with 
Green's functions under different flavours
\cite{bib:takayama76,bib:solovyev95,bib:ke15,bib:inoue15}.

In this work, we address various methodological aspects
of the application of DFT to the study of the MCA
arising from spin-orbit effects in the bandstructure of
extended systems using FeMe (Me$=$Co, Cu, Pd, Pt, and Au)
alloys with $L1_0$ structure as case studies (see Fig.~\ref{fig:l10}).
Methodologically, our aim is two-fold:
(i) focusing on the physics of the problem,
we take the SCF MAEs as reference values and examine the performance
of the NSCF and 2PT approximations, and 
(ii) on the computational side, we analyze the convergence
criteria required to get an accurate MAE with each method.
The chosen model systems allow to establish the
applicability of these approximations in terms of the SOI strength
(larger as we move downwards in the periodic table)
and of the character of the hybrid bands ($d-d$ for
FeCo, FePd, and FePt, and $d-s$ for FeCu and FeAu).

\begin{figure}
\includegraphics[width=\columnwidth]{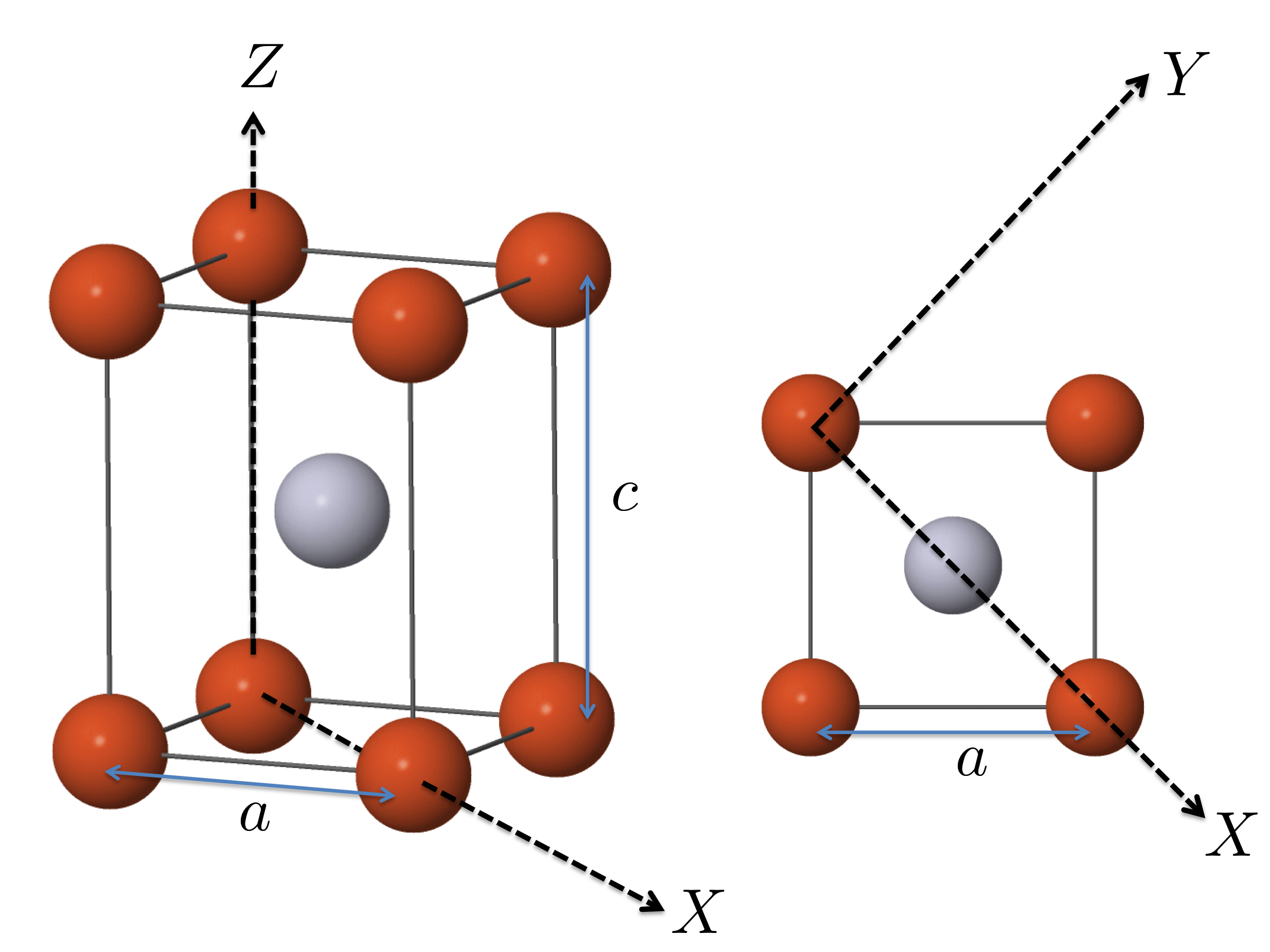}
\caption{\label{fig:l10}
Structure of the $L1_0$ tetragonal unit cell and cartesian axes
that provide a good match between the MLWFs and the $Y_{2m}$
orbital functions.}
\end{figure}

Our calculations show that the 2PT description of the MAE of the
lighter alloys is very reliable even under conditions
of strong bias voltage, while it
breaks down for heavier elements. In general, the use of 
SCF hardly changes the NSCF MAEs, as long as the values are
carefully converged.
Regarding the nature of the MCA in this family of alloys, we find
its physical origin in the availability of empty Fe electron states,
although the whole valence bandstructure contributes to its final magnitude.
The results presented here point to perturbative approaches as a feasible route
to modelling the MCA and related properties of magnetic metals,
specifically of tetragonal alloys, in working conditions under electrostatic fields.
The general character of these approaches suggests that they can be applied 
to lower-dimensional systems featuring dispersive bands.

The paper is organized as follows: section \ref{sec:theory} describes the
three methods used here to compute the MAE, namely SCF (section \ref{sec:scf}),
NSCF (section \ref{sec:nscf}) and, in more detail, two different implementations of
a 2PT formula in DFT codes (section \ref{sec:2pt}).
Details of the DFT calculation parameters for the Fe-based alloys
are presented in section \ref{sec:methods}.
The results for the charge neutral and non-neutral cases are
shown in sections \ref{sec:neutral} and \ref{sec:noneutral}, respectively.
Finally, conclusions are drawn in section \ref{sec:conclusions}.

\section{Theoretical background}
\label{sec:theory}

The spin-orbit interaction (SOI) Hamiltonian is generally
written as a sum over one-electron operators:
\begin{equation}
\label{eq:HSO}
H_{SO}=\sum_i \xi_i \mathbf{l}_i \cdot \mathbf{s}_i
\end{equation}
where $\mathbf{l}_i$ and $\mathbf{s}_i$ are the orbital and spin 
momentum operators, respectively, acting on the $i$-th 
electron in the system and $\xi_i$ is a constant 
that accounts for the SOI strength. 
In practice, as most of the relevant electronic and magnetic properties of 
solids derived from SOI originate from valence electrons, only outer-shell and 
semi-core electrons are considered in our first principles calculations.
Since $\xi_i$ is proportional to the radial
derivative of the potential, it increases with the atomic number. Furthermore, 
it is often a good approximation to take the same value for all the electrons
within the same $l$-shell.

In the next subsections we describe the three methods considered
in this work to evaluate the MAEs from first-principles, all of them including 
SOI at different levels of approximation. In particular, we will examine 
under which conditions a second-order perturbative approach, 
where $\xi_i$ acts as the perturbation constant, breaks down.

\subsection{Self-consistent MAE (SCF)}
\label{sec:scf}
%
Our reference \emph{ab initio} MAE value
is obtained by substracting the total energies $E_{tot}$
between two fully-relativistic self-consistent  
calculations, which include SOI, 
for two different orientations of the magnetization,
\begin{equation}
\label{eq:scf}
\mathrm{MAE} = E_{tot}^{x}-E_{tot}^{z}
\end{equation}
where the spins are aligned along the $OX$ and $OZ$ directions 
shown in the $L1_0$ unit cell model of Fig.~\ref{fig:l10}.
The main shortcoming of this method is its computational cost,
since Eq.~\ref{eq:scf} implies substracting two large numbers, 
which requires demanding convergence criteria. 
In fact, the obtention of well-converged MAEs 
from Eq.~\ref{eq:scf} is crucial in this work (see details in the
next section), since they 
are used as a benchmark for the approximations 
described in the next subsections.

\subsection{Non-self-consistent MAE (NSCF)}
\label{sec:nscf}

A scalar relativistic ground state (GS) is converged
in a spin-polarized calculation without SOI.
The so-obtained charge density is used to initialize a 
fully-relativistic calculation (i.e. non-collinear)
by turning it into a block-diagonal charge density matrix.
Then, the spin-orbit $H_{SO}$ term calculated for a given magnetization
axis is added to the scalar-relativistic hamiltonian $H_0$
and new eigenvalues are
calculated by diagonalization without further self-consistent cycles.
We denote the resulting total energy change $\Delta E_{tot}^{x,z}$ 
and the corresponding charge density change $\Delta \rho^{x,z}$.
The MAE is approximated as the difference in the band energies,
$E_{band}^{x,z}$, between the two orientations of the magnetization,
bearing in mind that the Fermi levels are in general different for 
the two orientations, as they are computed independently,
\begin{align}
\label{eq:kelly}
\mathrm{MAE}
   & \simeq  \Delta E_{tot}^{x} - \Delta E_{tot}^{z} 
    \simeq E_{band}^{x}-E_{band}^{z}  
   \nonumber \\
   & =  \sum_{k}^{N_k} \sum_{n}^{N_b}[ f^{x}(\epsilon_{k n}^{x}) \epsilon_{k n}^{x} -
               f^{z}(\epsilon_{k n}^{z}) \epsilon_{k n}^{z} ]
\end{align}
Here, the sum runs over one-electron eigenvalues $\epsilon_{k n}^{x,z}$, 
calculated with the spins aligned along the $OX$ and $OZ$ directions,
respectively, and integrated
over the entire first Brillouin Zone (1BZ). $N_b$ bands are considered (index $n$)
and a discrete grid of $N_k$ points (index $k$) is used to sample the 1BZ.
$f^{x,z}$ are the Fermi-Dirac distribution functions, which depend on the 
magnetization axes through the Fermi energy, while the finite electronic
temperature $kT$ acts as a smearing parameter.
The approximation is
based on the fact that $\Delta E_{tot}^x$ and $\Delta E_{tot}^z$ 
are correct to order $(\Delta \rho^x)^2$ and $(\Delta \rho^z)^2$, 
respectively. The method is thus sometimes called ``second variation'' \cite{bib:li90}
or ``force theorem'' \cite{bib:weinert85,bib:daalderop90}.
Eq.~\ref{eq:kelly} is correct to order $\Delta \rho^{x,z}$ while
the $(\Delta \rho^{x,z})^2$-order corrections have a small effect, since there 
are cancellations from the two magnetization directions \cite{bib:daalderop90}.
If one further assumes that the self-consistency cycles
introduce negligible modifications in the charge density matrix and in 
the exchange and correlation potential, then
Eq.~\ref{eq:scf} and Eq.~\ref{eq:kelly} should provide 
very similar MAE values, although with a considerable reduction in the
computational cost in the latter case.

\subsection{Second-order perturbative MAE (2PT)}
\label{sec:2pt}

A widely used alternative approximation treats the SOI
as a second-order perturbation (2PT) to the many-body GS,
$| \Psi^{(0)} \rangle$.
The general expression for the 2PT energy correction is 
\begin{equation}
\label{eq:de2general}
\Delta E_{SO}^{(2)} = \sum_{i \neq 0}
   \frac{\langle \Psi^{(0)}|H_{SO}|\Psi^{(i)}\rangle
         \langle \Psi^{(i)}|H_{SO}|\Psi^{(0)}\rangle}
        {E_{0}-E_{i}}
\end{equation}
where the sum runs over excited states. In a many-body language the
GS $| \Psi^{(0)} \rangle$ is formed by occupation of the lowest-lying
one-electron Kohn-Sham eigenstates up to the Fermi level. 
Each excited state $|\Psi^{(i)}\rangle$ is then constructed 
by creating electron-hole ($e-h$) pairs
using the unoccupied Kohn-Sham eigenstates.
Thus, the $E_0-E_i$ term in the denominator 
is simply the energy difference between the occupied and
unoccupied eigenvalues associated to the particular $e-h$ excitation
\cite{bib:takayama76,bib:bruno89,bib:cinal94,bib:vanderlaan98}. 
The perturbative expansion may then be written in terms
of the GS Kohn-Sham eigenstates $|kn\sigma\rangle$ as follows:
\begin{align}
\label{eq:de2kgrid}
\Delta E_{SO}^{(2)} & = 
    \frac{1}{N_k} \sum_{k}^{N_k} \sum_{n,n'}^{N_b} \sum_{\sigma,\sigma'}
    \frac{f(\epsilon_{kn\sigma})[1-f(\epsilon_{kn'\sigma'})]}
         {\epsilon_{kn\sigma}-\epsilon_{kn'\sigma'}}
   \; \times \nonumber\\
   & \langle kn\sigma   |H_{SO}| kn'\sigma' \rangle
     \langle kn'\sigma' |H_{SO}| kn\sigma   \rangle
\end{align}
where $\sigma,\sigma'$stand for the spin indices.

The second order formula given by Eq.~\ref{eq:de2kgrid} is applicable only to 
a non-degenerate ground state.
A degenerate ground state in an extended metallic system 
happens when there is a band crossing at a certain $k$-point 
precisely at the  Fermi level and this band pair is coupled by $H_{SO}$
(i.e. the corresponding matrix element is not zero) 
This can happen eventually, and in this situation the 
eigenstate pair should be treated separately by first-order 
degenerate state perturbation theory. 
However, in a calculation with a large $N_k$ 
the contribution to $\Delta E_{SO}^{(2)}$ 
of these exactly-degenerate states 
would be negligible, since only a handful 
of band crossings are expected at the Fermi level, and they
contribute with a factor of order $\xi / N_k$ \ \cite{bib:wang93}.
A sufficiently fine $k$-grid can map the spin-orbit 
band splitting effect nearby the crossing, 
so that  Eq.~\ref{eq:de2kgrid} can be safely used. 

In practice, it is convenient to express
Eq.~\ref{eq:de2kgrid} in a basis whose elements have well defined $lm$ quantum
numbers (spherical harmonics $Y_{lm}$).
A natural choice are atomic
orbitals (AOs), which already constitute the basis set in a number of DFT-based 
packages, leading to Bloch Kohn-Sham eigenstates of the form:
\begin{equation}
\label{eq:lcao}
|kn\sigma\rangle = \frac{1}{\sqrt{N_k}} \sum_{\mathbf R} e^{i{\mathbf{k}\cdot\mathbf{R}}}
\sum_\alpha^{N_\alpha} c^{n\sigma}_\alpha(k) |\alpha({\mathbf R}),\sigma\rangle
\end{equation}
where ${\mathbf R}$ runs over the lattice vectors, $|\alpha({\mathbf R})\rangle$
denotes an AO located in unit cell ${\mathbf R}$
and $N_\alpha$ is the total number of AOs in the basis set.
Inserting Eq.~\ref{eq:lcao} into \ref{eq:de2kgrid} yields:
\begin{align}
\label{eq:vdllcao}
\Delta E^{(2)}_{SO} & = \frac{1}{N_k} 
       \sum^{N_k}_k \sum_{\sigma\sigma'} 
       \sum^{N_\alpha}_{n,n'} 
    \frac{f(\epsilon_{kn\sigma})[1-f(\epsilon_{kn'\sigma'})]}
         {\epsilon_{kn\sigma}-\epsilon_{kn'\sigma'}} \; \times\nonumber \\
     & \sum^{N_\alpha}_{\alpha \beta,\alpha'\beta'} 
         \langle \alpha |H^{\sigma\sigma'}_{SO}(k)|\beta \rangle
         \langle \alpha'|H^{\sigma'\sigma}_{SO}(k)|\beta'\rangle
        n^{nn'\sigma\sigma'}_{\alpha\beta}(k)
        n^{n'n\sigma'\sigma}_{\alpha'\beta'}(k)
\end{align}
where the k-space $H_{SO}(k)$ matrix elements are given by:
\begin{equation}
\label{eq:hkso}
\langle \alpha|H^{\sigma\sigma'}_{SO}(k)|\beta\rangle= \frac{1}{N_k}
  \sum_{\mathbf{R}} e^{i\mathbf{k}\cdot\mathbf{R}} 
  \langle \alpha({\mathbf 0}),\sigma|H_{SO}|\beta({\mathbf R}),\sigma'\rangle
\end{equation}
and the {\it generalized projected charges} by:
\begin{equation}
\label{eq:nn}
 n^{nn'\sigma\sigma'}_{\alpha\beta} (k) = (c^{n\sigma}_\alpha (k))^*
                                       c^{n'\sigma'}_\beta (k)
\end{equation}

It is usual to further assume the so-called on-site approximation, whereby
the $\mathbf{l}_i \cdot \mathbf{s}_i$ operators only mix states within the
same $l$-shell of a given atom contained in the origin unit cell. 
Eq.~\ref{eq:hkso} then becomes:
\begin{align}
\label{eq:onsite}
\langle \alpha|H^{\sigma\sigma'}_{SO}(k)|\beta\rangle & \approx 
  \langle \alpha({\mathbf 0}),\sigma|H_{SO}|\beta({\mathbf 0}),\sigma'\rangle
  \nonumber \\
  & \approx 
  \xi_{\alpha,l}
   \langle \alpha lm \sigma | {\mathbf{ l\cdot s}} | \beta l'm' \sigma' \rangle
    \delta_{\alpha\beta} \delta_{ll'}
\end{align}
where indices $\alpha,\beta$ in the last term now refer to the principal quantum
number in a given atom (in a multi-$\zeta$ scheme, also the particular $\zeta$),
and $lm$ stand for the orbital and magnetic quantum numbers of each AO.
$\xi_{\alpha,l}$ is the SOI strength for this $l$-shell resulting from the 
integration of the radial part in the 
$\langle \alpha lm \sigma | H_{SO} | \alpha lm' \sigma' \rangle$ 
matrix elements (independent of $m m'$ and $\sigma \sigma'$). 
The angular part of these matrix elements, 
$\langle \alpha lm \sigma | \mathbf{l} \cdot \mathbf{s} |\alpha lm' \sigma'\rangle$,
take simple analytical expressions and tabulated formulas as a function
of the spherical harmonics $Y_{lm}$ involved can be found, for example,
in Refs.~\cite{bib:abate65,bib:elsasser88}.
The on-site approximation simplifies considerably the 2PT formula:
\begin{align}
\label{eq:vdllcao_onsite}
\Delta E^{(2)}_{SO} & = \frac{1}{N_k}
       \sum^{N_k}_k \sum_{\sigma\sigma'}
       \sum^{N_\alpha}_{n,n'}
    \frac{f(\epsilon_{kn\sigma})[1-f(\epsilon_{kn'\sigma'})]}
         {\epsilon_{kn\sigma}-\epsilon_{kn'\sigma'}} \; \times \nonumber\\
      &  A^{nn'\sigma\sigma'}(k) A^{n'n\sigma'\sigma}(k)
\end{align}
where we have defined:
\begin{equation}
 A^{nn'\sigma\sigma'}(k) = \sum_{\alpha l} \xi_{\alpha l} 
   \sum_{mm'}^{2l+1} \langle \alpha lm \sigma |{\mathbf{l\cdot s}}|\alpha lm' \sigma' \rangle
     n^{nn'\sigma\sigma'}_{\alpha lm \alpha lm'}(k)
\end{equation}
We observe that the 2PT Eqs.~\ref{eq:vdllcao} and \ref{eq:vdllcao_onsite}
are perturbative in the SOI strength, which appears explicitly in the form 
of the parameters $\xi_{\alpha l}$ in this equation.

Next we consider the case when the unperturbed spin-polarized
calculation is realized employing a plane-wave basis set, as many DFT
codes do.
Instead of using a Bloch-function representation of the 
the Kohn-Sham eigenstates, we use a set of $N_w$ 
maximally localized Wannier functions (MLWF) as 
formulated in Ref.~\cite{bib:marzari97}.
MLWFs are constructed to yield the exact 
eigenvalues as the {\it ab initio} calculation.
Usually, a previous disentanglement procedure is 
carried out, whereby a handful of relevant bands within 
an energy window are isolated from the rest \cite{bib:souza01}. 
For the systems under study, we focus on the bands that 
originate from the $d$-valence electrons of both metal atoms
(allowing also for some degree of $s-d$ hybridization)
and belong to a window of about 10~eV below and 5~eV above
the Fermi energy.
Afterwards, it is straightforward to obtain new Bloch functions 
on a $k$-grid as dense as desired by interpolation \cite{bib:yates07}. 
This procedure allows us to estimate the 2PT MAE using as input
solely a scalar-relativistic first-principles calculation and 
does not require a highly dense 1BZ $k$-sampling.

The $j$-th MLWF localized at the the unit cell 
$\mathbf{R}$ that results from band 
disentanglement and wannierization for 
states with spin $\sigma$ is:
\begin{equation}
\label{eq:wannier}
|w_{j}^{\sigma} (\mathbf{R}) \rangle = \frac{1}{N_k}
    \sum^{N_k}_k e^{-i\mathbf{k} \cdot \mathbf{R}}
    \sum_n^{N_b} Q_{j}^{n \sigma}(k) |k n \sigma \rangle
\end{equation}
where $|k n \sigma \rangle$ stands for 
the Kohn-Sham eigenstates already interpolated in the dense
$k$-grid \cite{bib:yates07}
and $Q_{j}^{n \sigma}(\mathbf{k})$ 
are the coefficients that relate the Wannier and Bloch functions.

Typically, atomic-like wavefunctions, formed by a radial function and  
spherical harmonics $Y_{lm}$ to describe the angular component,
are used to initialize the wannierization procedure. 
Here, we use $d$-orbitals centred at the atomic sites
and a few $s$-waves at interstitial positions. 
The purpose of the latter functions is to facilitate the wannierization 
and will not take part in the 2PT MAE calculation. 
We fix $l=2$ and drop this index in the following. 
Thus, the $j$ label accounts for the $\alpha$-th atom in the cell $\mathbf{R}$
and the $m=0, \pm 1, \pm 2$ quantum number.
If the deviation of the MLWFs from actual atomic
wavefunctions is small, we can approximate the matrix elements
$\langle w_{\alpha m}^{\sigma} (\mathbf{R}) | H_{SO} | w_{\alpha m'}^{\sigma'} (\mathbf{R}) \rangle$ 
by the ones in the AO representation 
$\langle \alpha m \sigma | \mathbf{l} \cdot \mathbf{s} |\alpha m' \sigma' \rangle$
and take advantage of their simple analytic expressions \cite{bib:abate65,bib:elsasser88}.
Note that, in general, the resulting MLWFs do not keep a well-defined orbital character
because they have to account for both the intra- and inter-atomic orbital
hybridization present in the system \cite{bib:roychoudhury17}.
Nevertheless, a suitable choice of axes in the systems under study 
allows to obtain MLWFs that keep the atomic orbital character 
and justify this approximation.
Substituting Eq.~\ref{eq:wannier} in Eq.~\ref{eq:de2kgrid} and using this 
approach, the second order energy correction associated to the SOI is
\begin{align}
\label{eq:vdl}
\Delta E^{(2)}_{SO} & = 
       \sum_{\alpha_1 \alpha_2} \sum_{\sigma\sigma'}
       \sum_{m_1m_2} \sum_{m'_1 m'_2}
       \xi_{\alpha_1} \xi_{\alpha_2}
       F^{\sigma\sigma'\alpha_1 \alpha_2}_{m_1 m_2 m'_1 m'_2} \; \times \nonumber\\
    &  \langle \alpha_1 m'_1 \sigma' | \mathbf{l}\cdot\mathbf{s} | \alpha_1 m_1 \sigma \rangle
       \langle \alpha_2 m'_2 \sigma' | \mathbf{l}\cdot\mathbf{s} | \alpha_2 m_2 \sigma \rangle
\end{align}
where the $m_i$ indexes run over the 5 $d$-orbitals of each atom $\alpha_i$ in the unit cell.
The $F$ coefficients contain the details of the basis change
from Kohn-Sham states to MLWF states and, implicitly,
they allow us to use a dense $k$-grid by interpolation:
\begin{align}
\label{eq:vdlcoefs}
F^{\sigma \sigma' \alpha_1 \alpha_2}_{m_1 m_2 m'_1 m'_2} & = 
   \frac{1}{N_k} \sum^{N_k}_k \sum_{nn'}^{N_w}
   \frac{ f(\epsilon_{kn\sigma}) [1-f(\epsilon_{kn'\sigma'})]}
   {\epsilon_{kn\sigma}-\epsilon_{kn'\sigma'}} \; \times \nonumber\\
 & Q^{n' \sigma'}_{\alpha_1 m'_1}(k)
   (Q^{n \sigma}_{\alpha_1 m_1}(k))^{\star}
   Q^{n \sigma}_{\alpha_2 m_2}(k)
   (Q^{n' \sigma'}_{\alpha_2 m'_2}(k))^{\star}
\end{align}
Eq.~\ref{eq:vdl} is similar to the 2PT MAE formula 
for extended systems developed by van der Laan 
\cite{bib:vanderlaan98}, 
and here we generalize the result to the case of more than one atom per unit cell. 
It is also straigthforward to show that
the on-site approximation Eq.~\ref{eq:vdllcao_onsite} reduces to the above
expression after
replacing the Bloch coefficients $c_\alpha^{n\sigma}(k)$ in Eq.~\ref{eq:nn} by the
$Q_j^{n\sigma}(k)$ ones and restricting the summation over the angular
momentum numbers to the $l=l'=2$ case.

Note that Eqs.~\ref{eq:vdl} (or Eq.~\ref{eq:vdllcao}) is a ``four-legged'' expression
in the sense that is contributed by two different $e-h$ pairs.
There are other 2PT formulations based on the use of a localized basis set of orbitals 
\cite{bib:bruno89,bib:cinal94,bib:takayama76,bib:ke15}.
However, it is worth to mention that our formulation of the MAE 
does not neglect 
spin-flip contributions, unlike the one proposed by 
Bruno \cite{bib:bruno89}. Formulas that neglect wavefunction phase effects 
in the Kohn-Sham-to-local-basis projection  
have also been proposed \cite{bib:takayama76,bib:ke15}.
By doing so, the ``four-legged'' Eq.~\ref{eq:vdl} becomes 
only ``two-legged'' and Eq.~\ref{eq:vdlcoefs} can be written in simpler
terms, namely the projected densities of states on the local orbitals.

\section{Computational details}
\label{sec:methods}

In all the procedures described in this section,
to prevent the MAE values from being biased by the structural
parameters, we have kept the lattice constants $a,c$
fixed in all calculations (see Fig.~\ref{fig:l10}
and Table~\ref{tab:VASPWANNparameters}).
The model for FeCo has the Fe $bcc$ unit cell volume and $c/a=1.2$
to maximise the anisotropy, as suggested by the literature on
strain effects \cite{bib:burkert04}.
For FeCu, we keep the Cu-Cu as in bulk $fcc$ Cu and $c/a=1.34$ \cite{bib:errandonea97}.
Finally, we use published lattice constants for FePd, FeAu \cite{bib:galanakis00} and
FePt \cite{bib:khan16}.

\begin{table*}
\caption{\label{tab:VASPWANNparameters}
Unit cell lattice parameters (see Fig.~\ref{fig:l10}) of the
considered model systems and calculation parameters
used in the wannierization.
$n_s$ is the number of $s$-wave-like Wannier functions, introduced
in addition to the $d$-orbital-like ones, placed at interstitial
sites of the structure.
$[w_0,w_1]$ are the frozen windows used for
disentanglement with respect to the Fermi energy.
Two intervals are shown for FeAu that correspond to different
windows for the spin-majority and spin-minority bands, respectively.
${k}_1$ is the grid used in the reference DFT calculation and
${k}_2$ the interpolated grid used to evaluate the MAE in the 2PT
approximation. $p_{min}$ is the minimum projection factor
$p_{\alpha m \sigma}$ (Eq.~\ref{eq:orbiwann}) found for each system.
}
\begin{ruledtabular}
\begin{tabular}{cccccc}
          &   FeCo  & FeCu  & FePd  & FePt & FeAu \\
$a$ (\AA) &  2.680  & 2.553 & 2.751 & 2.722 & 2.885 \\
$c/a$     &  1.2    & 1.339 & 1.327 & 1.364 & 1.328 \\
$n_s$     &  3 & 4 & 3 & 2 & 3,4 \\
$[w_0,w_1]$ (eV) & $[-5.2,2.8]$ & $[-10.0,3.5]$ & $[-6.6,2.4]$ & $[-5.8,2.2]$ & $[-3.4,-1.1],[-1.4,1.6]$ \\
${k}_1$   &  $16 \times 16 \times 14$ & $16 \times 16 \times 12$ & $16 \times 16 \times 12$ & $16 \times 16 \times 12$ & $16 \times 16 \times 12$ \\
${k}_2$   &  $40 \times 40 \times 33$ & $36 \times 36 \times 28$ &  $36 \times 36 \times 27$ &  $40 \times 40 \times 30$ &  $36 \times 36 \times 27$ \\
$p_{min}$ & 0.88 & 0.82 & 0.82 & 0.90 & 0.85 \\
\end{tabular}
\end{ruledtabular}
\end{table*}

Two DFT codes have been used in this work:
SIESTA-Green~\cite{bib:soler02,bib:cuadrado12} (SG)
and  VASP \cite{bib:kresse93,bib:kresse96}.
The former uses multi-$\zeta$ non-orthogonal strictly localized 
numerical AOs as basis set and replaces core electrons by pseudo-potentials,
while
the latter uses plane-waves and projector-augmented 
wave potentials to describe the ion cores \cite{bib:bloechl94}.
Both codes feature SOI implementations \cite{bib:cuadrado12,bib:steiner16} 
that allow to obtain the MAE values directly from Eq.~\ref{eq:scf} 
or the SOI-corrected eigenvalues of  Eq.~\ref{eq:kelly}.
By working with both codes we ensure that the 
conclusions about the MCA are not biased by the basis set type.
The exchange and correlation functional used in all calculations is
that of Perdew, Burke and Ernzerhof (PBE) \cite{bib:pbe96}.

In the VASP calculations the $p$ semi-core states are included as valence electrons.
We have set the plane-wave energy cut-off to
400\,eV in all the systems, except for 420\,eV in FeAu,
and suitable $k$-point Monkhorst-Pack grids \cite{bib:monk76}
according to each lattice dimensions and calculation type
[($24 \times 24 \times 20$) for FeCo and
($24 \times 24 \times 18$) for others]. 
The tetrahedron method with Bl\"ochl corrections is used to obtain
the Fermi level \cite{bib:bloechl94b}.
For the SG calculations we have adopted a double-$\zeta$ polarized
scheme to generate the AO basis set using a confinement energy of 100~meV
although, as opposed to VASP, no $p$ semi-core states are considered.
Pseudo-core corrections are included for all the atoms involved, while
a very fine real space mesh is employed by setting the mesh cut-off to 4000~Ry. 

In the SG case, SOI matrices are calculated under the fully-relativistic pseudo-potential
(FR-PP) method described elsewhere~\cite{bib:cuadrado12}. This approach
goes beyond the on-site approximation~\cite{bib:fernandez06} 
(Eq.~\ref{eq:onsite})
as it takes into account intra- and inter-atomic interactions between different
$l$-shells. Although the off-site terms tend to be small, test calculations
show that neglecting them can induce errors in the MAE of around 0.2~meV or
even larger (around 1~meV) in particular cases. Nevertheless, the on-site
approximation allows to extract the SOI strengths $\xi_{\alpha l}$, which
we provide in Table~\ref{tab:muSL} for the $d$ orbitals.

\begin{table*}
\caption{\label{tab:muSL}
SG values of the atomic SOI strength parameter 
$\xi_\mathrm{Me}$ (Me=Co,Cu,Pd,Pt,Au) and
VASP values of the atomic spin and orbital magnetic moments,
$\mu_S$ and $\mu_L$, respectively, in Bohr magnetons 
($\mu_S$ values are obtained from calculations without SOI).
The last column shows an approximated MAE obtained from the expression
$-\sum_{\alpha} \frac{\xi_\alpha}{4} [ \mu_L^x(\alpha) - \mu_L^z(\alpha) ]$,
where the index $\alpha$ runs over the Fe and Me atoms, with
$\xi_{\mathrm{Fe}}=59.65$\,meV.}
\begin{ruledtabular}
\begin{tabular}{ccccccccc}
    & $\xi_\mathrm{Me}$ (meV) 
    & $\mu_S$(Fe) & $\mu_S$(Me) 
    & $\mu_L^x$(Fe) & $\mu_L^x$(Me) 
    & $\mu_L^z$(Fe) & $\mu_L^z$(Me) & MAE (meV) \\
FeCo &  74.12 & 2.74 & 1.67 & 0.053 & 0.077 & 0.064 & 0.088 & 0.37  \\
FeCu & 110.44 & 2.55 & 0.16 & 0.045 & 0.009 & 0.055 & 0.011 & 0.20  \\
FePd & 191.36 & 3.00 & 0.38 & 0.061 & 0.030 & 0.069 & 0.027 & -0.02 \\
FePt & 537.18 & 2.93 & 0.40 & 0.057 & 0.056 & 0.060 & 0.044 & -1.57 \\
FeAu & 615.05 & 2.98 & 0.03 & 0.045 & 0.037 & 0.065 & 0.029 & -0.93 \\
\end{tabular}
\end{ruledtabular}
\end{table*}

Even when working with the SCF method for SOI,
the calculation parameters must be carefully tested.
The Fermi energy smearing is a decisive technical factor for the MAE of
some of the systems. This issue has been addressed with both codes for the
SCF and NSCF methods.
We find satisfactory convergence when the
tetrahedron method is used for the SCF MAE with VASP \cite{bib:bloechl94b}.
By doing so, we avoid $kT$-dependence in the
resulting MAE values (we have checked that the total energy
extrapolation to $kT=0$ gives, in general,
good agreement with the results of the tetrahedron method).
With SG, since the use of finer $k$-point grids
can be afforded at a reasonable computational and memory cost,
a high convergence could be systematically achieved
in the $k$-grid. Convergency values below 0.01~meV are obtained with $kT$ 
values entering the Fermi-Dirac distribution function as low as 1~meV. 
In the NSCF calculations we find that the smearing function, whether Fermi-Dirac
or Methfessel-Paxton of a given order \cite{bib:methfessel89},
has a non-negligible influence, but it becomes less important for
sufficiently fine $k$-grids and small $kT$.
A detailed convergence analysis for all phases 
can be found in the Supplementary Information (see tables in sections I,II
and Figs.~1 and 2).

For the more elaborate 2PT methodology, we have implemented 
Eq.~\ref{eq:vdllcao} for the SG calculations and
the semi-analytical form of Eq.~\ref{eq:vdl} for VASP in combination with 
Wannier90 \cite{bib:mostofi14,bib:marzari97}.  
The $k$-grids needed to obtain a 
faithful representation of the electronic structure with MLWFs can be less 
dense than the ones typically needed to obtain the MAEs. The latter, also used
in the explicit evaluation of Eqs.~\ref{eq:vdl} and \ref{eq:vdlcoefs},
can be chosen as dense as desired by MLWF interpolation \cite{bib:yates07}. 
Besides, due to the strongly hybridized $d$-bands in the $L1_0$ Fe-alloys, 
prior to wannierization it is useful to perform a disentanglement of the 
bands \cite{bib:souza01} within an energy window that contains the relevant states. 
A numerical drawback in the whole procedure is that the quality of the wannierization is
system-dependent. Therefore, for each alloy we have chosen suitable settings, 
shown in Table~\ref{tab:VASPWANNparameters}, to meet the usual sanity
requirements of a wannierization, namely, little overlap between MLWFs (at least between the 
functions that emulate the $d$-orbitals), bandstructure reproducibility, and 
spatial localization. 

It is worth to mention that the five atomic $d$-orbital wavefunction geometries, i.e.
the representations of the angular functions $Y_{2m} (\hat r)$ in cartesian coordinates,
depend on the convention taken for the $OX,OY,OZ$ axes in each code.
Obviously, the electronic structure that results from hybridization of the 
atomic wavefunctions must be independent of those conventions.
However, if we align the crystallographic directions and the cartesian axes of 
VASP and Wannier90 as shown in Fig.~\ref{fig:l10},
we can represent the bonding states as overlapping 
$Y_{2m} (\widehat{\mathbf{r}-\mathbf{R}_\alpha})$
functions localized at atomic positions $\mathbf{R}_\alpha$. Otherwise, the
overlaps would happen for linear combinations of $Y_{2m} (\hat r)$ functions
at each site. In such case, the MLWFs that reproduce the electronic structure
do not resemble the initial spherical harmonics and Eq.~\ref{eq:vdl}
cannot be applied directly: an intermediate step would be
needed to account for the linear combinations of $Y_{2m} (\hat r)$ functions.
The axes choice of Fig.~\ref{fig:l10} makes this step unnecessary.
Indeed, we can trace the MLWFs back to individual Fe and Me $d$-orbitals
by visual inspection (see Supplementary Information Fig.~3), 
deviations being the result of the inter-atomic hybridization only,
i.e. an effect of a purely physical origin, and not an spurious one caused
by the axes convention.
To quantify those deviations, we define the projections 
\begin{equation}
\label{eq:orbiwann}
p_{\alpha m \sigma} = | \langle w_{\alpha m}^{\sigma} (\mathbf{R}) |  \alpha m \sigma \rangle |^2
\end{equation}
which range between zero and one.
As shown in Table~\ref{tab:VASPWANNparameters}, 
despite the dispersive character of the bandstructures, 
in this work we find projections above 0.80 after wannierization.

\section{Results and discussion}
\label{sec:results}

\subsection{Neutral systems}
\label{sec:neutral}

Table~\ref{tab:mae} contains the collection of the MAE values
calculated with the methodologies presented in Section~\ref{sec:methods}.
All the approaches provide the same behaviour in the
MCA of each alloy, albeit there are some quantitative
differences in the corresponding MAE values, which will be discussed below.
The easy magnetization axis is $OZ$ in the five studied systems.
Overall, MAE values are smaller than 0.5\,meV except for FePt,
a well-known prototype of large
MCA, which shows a MAE of nearly 3\,meV
in good agreement with other {\it ab-initio} values available in the 
literature \cite{bib:galanakis00,bib:burkert05,bib:cuadrado12,bib:khan16}.

\begin{table*}
\caption{\label{tab:mae} MAE values in meV for the neutral systems.
Labels SG and V indicate SIESTA-Green and VASP calculations, respectively,
with the cut-off energy and $k$-point samplings discussed in the text.
In the 2PT SG (V)  calculations a Fermi-Dirac smearing with $kT=10$\,meV
($kT=50$\,meV) has been used. }
\begin{ruledtabular}
\begin{tabular}{ccccccc}
     &  SCF (SG) & SCF (V) & NSCF (SG) & NSCF (V) & 2PT (SG) & 2PT (V) \\
FeCo &  0.42  &  0.39  &  0.45  &  0.55  &  0.44  &  0.35  \\
FeCu &  0.38  &  0.45  &  0.42  &  0.45  &  0.42  &  0.38  \\
FePd &  0.22  &  0.16  &  0.20  &  0.13  &  0.19  & -0.11  \\
FePt &  2.93  &  2.59  &  2.93  &  2.78  &  2.92  &  0.63  \\
FeAu &  0.20  &  0.50  &  0.36  &  0.62  &  0.41  &  0.76  \\
\end{tabular}
\end{ruledtabular}
\end{table*}

The first important observation is that the MAE of Fe-based $L1_0$ alloys
is not directly correlated with the SO strength of the alloying metal. 
Indeed, it is the electronic structure
what governs the MCA of these alloys, overruling the effect of the SO strength.
We find larger MAE values for FeCo and FeCu than for FePd and FeAu
in despite of $\xi_{\mathrm{Pd,Au}}$ being larger than $\xi_{\mathrm{Co,Cu}}$
(see Table~\ref{tab:muSL}).
In the case of FeCo it is known that a large MAE can be achieved
by a strain on the cell along $OZ$. For $c/a=1.2$, the geometry
chosen for this work, a maximum is obtained because degenerate states
coupled by $H_{SO}$ lie on the Fermi level \cite{bib:burkert04}.
As a matter of fact, FeCo is not an isolated case \cite{bib:gambardella03,bib:ke15}.

The SCF values obtained in the VASP and SG calculations
are in good agreement for FeCo, FeCu, and FePd, where discrepancies
of 0.07~meV or smaller are found. For FePt and FeAu larger
deviations of around 0.3~meV exist.
The reason for this discrepancy is difficult to
identify. Small quantitative differences in the band structures provided
by both codes would be unimportant for most physical properties but,
unfortunately, they become significant for the estimation of the MAEs.
First, we note that the VASP values could not be converged in $k$-grid
at the same level of accuracy as the SG MAEs. Another source of
error could be associated to limitations of the SG calculations,
such as the basis set size and AO extensions, the neglect of semi-core $p$ 
states or even the actual FR-PP approximation.

The MAE values predicted by the NSCF method (see Table~\ref{tab:mae}) are, 
in general, very close to the SCF values, with typical
deviations well below 0.1~meV. 
It is striking to find such a good agreement
even for the heaviest metal systems,
where the charge density change is expected to be larger.
There are only a few exceptions,
namely the VASP calculation for FePt, FeCo and FeAu 
and the SG one for FeAu for which, nonetheless, differences remain smaller than 
0.2~meV. For the formers, we assign the discrepancy to
calculation details rather than to
the limitation of using the non-self-consistent charge density.
Among the technical details, it is the smearing method of the Fermi level
the crucial one, since the NSCF approach is sensitive to the
small Fermi energy shift when magnetization occurs in one or other direction.
In the FeAu case with SG, where full $k$-convergence is
achieved, we may consider the 0.12~meV deviation as an upper limit to the
accuracy of the NSCF, probably due to the larger $\xi_{{\mathrm Au}}$ SOI
strength.
Notwithstanding, the $H_{SO}$ term is fully accounted for by this approach.

We next address the performance of the 2PT approximation to the MAE.
If we first focus on the SG MAEs in Table~\ref{tab:mae}, they are in 
almost perfect agreement with the SCF and NSCF MAE values in all cases except, again,
the FeAu alloy which results 0.2~meV larger than the SCF value following
a similar trend as under the NSCF. Thus, the
same behaviour between the NSCF and 2PT methods indicates that their
accuracy is very similar despite the neglect of high-order
$H_{SO}$ terms in the latter, both providing excellent results at a much lower
computational cost compared to the reference SCF calculations.

The 2PT-derived MAE is determined by eigenstates
around the Fermi level within an energy range given by the SOI strength $\xi$ 
and the $(\epsilon_{kn\sigma}-\epsilon_{kn'\sigma'})^{-1}$ factors.
Indeed, Eq.~\ref{eq:vdllcao} gives less weight to the $e-h$ pairs
coupled by $H_{SO}$ matrix elements that lie farther in energy
(the contributions of the energy prefactors are shown in Supplementary Information Fig.~4).
Intuitively, deep states in the valence band might be regarded as negligible 
for the MAE. However, a third factor needs to be considered: the 
avaible number of states. To analyze the competition of these three effects,
we have calculated the 2PT MAE 
allowing only initial (final) states within an energy window
below (above) the Fermi energy when evaluating 
Eq.~\ref{eq:vdllcao}. The results are shown 
in Fig.~\ref{fig:fermiwinsiesta} for the SG case, 
while those with VASP are qualitatively the same (not shown). 
When imposing a window on the
occupied states, a plateau in the MAE value is not reached
until 5-7\,eV, depending on the system. 
These energy ranges comprise the
$d$-band widths below the Fermi level 
(see the densities of states (DOS) projected on the $d$-orbitals in 
the Supplementary Information Fig.~5).
With heavier atoms, deeper states contribute non-trivially
to the MAE, even reversing its sign, as it is the case of FeAu.
Therefore, the final value of the MAE of each system depends on 
its electronic structure details, since the dispersion and binding
energies of the individual band pairs coupled by $H_{SO}$ largely
vary from one system to another.
The effect of constraining the accessible empty states in
Fig.~\ref{fig:fermiwinsiesta} is less dramatic and reveals a
common behaviour in all the systems. With a window above the
Fermi level, the MAE plateau is reached at $\sim 2.5$~eV.
As shown in the DOS plots in the Supplementary Information (Fig.~5), 
this energy range corresponds to the empty spin-minority states of Fe
in all the alloys, and it is also contributed to a lesser extent
by the other metal empty $d$-states if available (Co and Pt).

\begin{figure}
\includegraphics[width=\columnwidth]{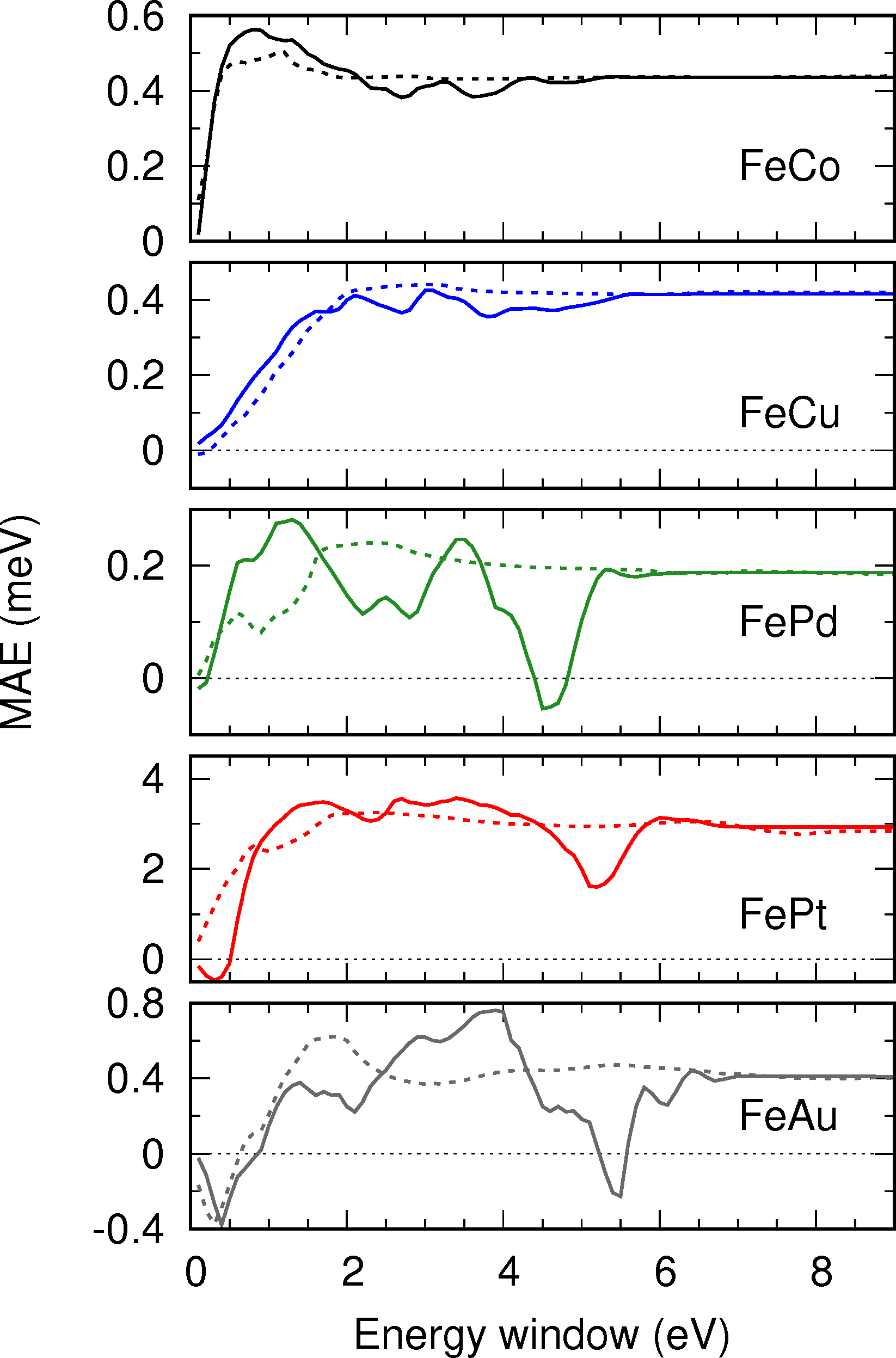}
\caption{\label{fig:fermiwinsiesta}
Dependence of the MAE calculated in the 2PT approximation with
Eq.~\ref{eq:vdllcao} (SG calculation) on the energy window of allowed
occupied (solid) and unoccupied (dashed) one-electron states.
}
\end{figure}

It is a general trend in these systems that the high DOS
counterbalances the decay of the $(\epsilon_{kn\sigma}-\epsilon_{kn'\sigma'})^{-1}$ factors.
In brief, Fig.~\ref{fig:fermiwinsiesta} shows that states distant
from the Fermi level by as much as several eV
(that is, spanning the whole $d$-band of the alloy) cannot be neglected
in a 2PT calculation, not even in the case of atoms with weak SOI.
We conclude that the accessibility of empty Fe spin-minority states is a common
feature that allows for sizeable MCAs in the Fe-based alloys,
while the details of the occupied $d$-bands of each case determine
the final MAE value. 

The 2PT MAE values obtained with the wannierized bands from the VASP
calculation yield worse agreement than the other methods 
becoming more pronounced the heavier the metal in the Fe alloy 
(see Table~\ref{tab:mae}).
Specifically, the easy axis for FePd is switched to $OX$ while the MAE 
of FePt is considerably underestimated and that of FeAu overestimated.
Although this method allows the use of very dense $k$-point grids
by interpolation, the quality of the wannierization procedure is
crucial for numerical accuracy. Nonetheless, the main factor that
undermines the final MAE values is the approximation in the
$H_{SO}$ matrix elements: the assumption that the MLWFs
correspond to atomic orbitals with well-defined $lm$ quantum numbers and,
to a lesser extent, the on-site approximation
(Eq.~\ref{eq:onsite}) and the neglect of $sp$-orbital contributions.
All in all Eq.~\ref{eq:vdl} is a coarse approximation.
Despite the apparent resemblance of MLWFs with atomic wavefunctions
by visual inspection (see Supplementary Information Fig.~3), the deformations of 
these ``$d$-orbitals'' are significant, due to the fact that Fe-alloys have
strongly-hybridized bands. 
For FeCo and FeCu, nevertheless, the MAE behaviour in the
energy windows analyzed is similar to the SG ones (not shown).

Another appealing aspect of the 2PT formulation is that it
allows  to split the MAE into contributions arising from pairs of atoms
in the metallic alloy: Fe-Fe, Fe-Me and Me-Me. 
This is straighforward when using MLWFs and Eq.~\ref{eq:vdl} 
(see Supplementary Information Fig.~6), while if
Eq.~\ref{eq:vdllcao} is used, the decomposition 
can be realized by restricting $(\alpha,\alpha')$  to a given pair 
of atoms and performing the summation over all the other 
$(\beta,\beta')$ AO contributions.
The result is shown in Fig.~\ref{fig:barplotSG}. 
As expected, Fe and Co have a balanced weight
on the final FeCo MAE, while Cu hardly contributes to that of FeCu.
In the rest of alloys
the decompositions show the contribution we could expect from the
knowledge of the electronic structure, at least qualitatively.
The Pt-Pt and Fe-Pt terms are the main contributors in the FePt alloy, 
because $\xi_{\mathrm{Pt}}$ is an order of magnitude 
larger than $\xi_{\mathrm{Fe}}$ and because the $d$ electrons of
both species are strongly hybridized.
We find, in agreement to Ref.~\cite{bib:solovyev95},
that the large contribution of the SOI of Pt atoms to the 
perpendicular anisotropy ($OZ$ easy axis) is partially balanced by the effect of the 
Fe-Pt hybrid bands, which favour in-plane anisotropy.
In FeAu, the Au-Au term is also very large 
due to the magnitude of $\xi_{\mathrm{Au}}$, but now 
we see that the contribution of Fe-Au terms is weaker, since 
there is little hybridization with Au-$d$ electrons.

\begin{figure}
\includegraphics[width=0.5\columnwidth]{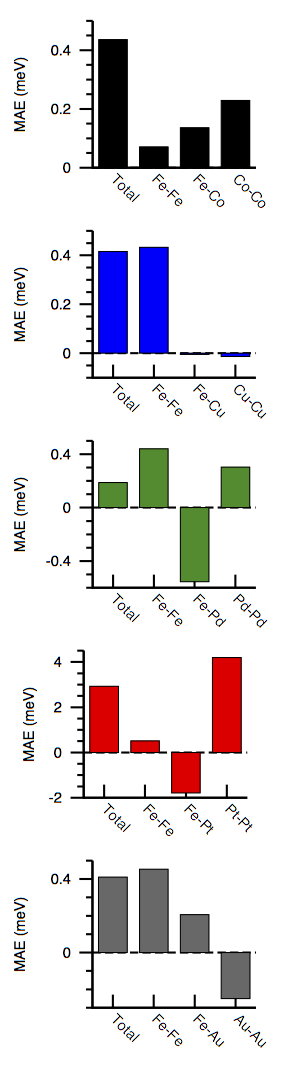}
\caption{\label{fig:barplotSG}
Contributions of the atom-pair terms of Eq.~\ref{eq:vdllcao}
(SG calculation)  to the 2PT MAE.}
\end{figure}

Finally, we compare our perturbative results 
with the widely used 2PT equation by Bruno for 
bands of $d$-orbital character \cite{bib:bruno89}, which
assumes that all the accesible holes are minority spin states. 
Therefore, it neglects $e-h$ excitations that involve spin-flip, which leads to
$\Delta E^{(2)} \propto \xi \langle \mathbf{L} \rangle$.
Table~\ref{tab:muSL} shows the MAE values obtained in this approximation with the
DFT-calculated atomic orbital moment values $\mu_L$ for each magnetization direction.
Although the density of majority-spin states above the Fermi level is marginal 
(shown in the Supplementary Information Fig.~5),
the results of the Bruno formula differ significantly from the 
other methods, and in some cases it does not even predict the correct easy 
magnetization axis. In addition to the breakdown of the 
approximation itself, we can ascribe the discrepancies to the 
tendency of DFT  to underestimate orbital moment values.

\subsection{Non-neutral systems}
\label{sec:noneutral}

Figure~\ref{fig:kelly_vdl} shows
the MAE as a function of the number of electrons $N_e$ calculated with
the NSCF and the 2PT approaches. 
Here, we have followed a rigid band
approach, in which the Fermi level is recalculated for each $N_e$ employing
the eigenvalues and eigenstates of the unperturbed neutral calculation.
Hence, the plots represent the MAE behaviour of each alloy under
conditions of doping or application of a bias voltage, which are
common working conditions in magnetic devices. Note, however, that due
to the rigid band approach
only small deviations from the neutral situation are physically meaningful.
As expected from the disccusion in the previous section, there are 
only small differences between the NSCF curves calculated with SG and VASP.

\begin{figure*}
\includegraphics[width=2.0\columnwidth]{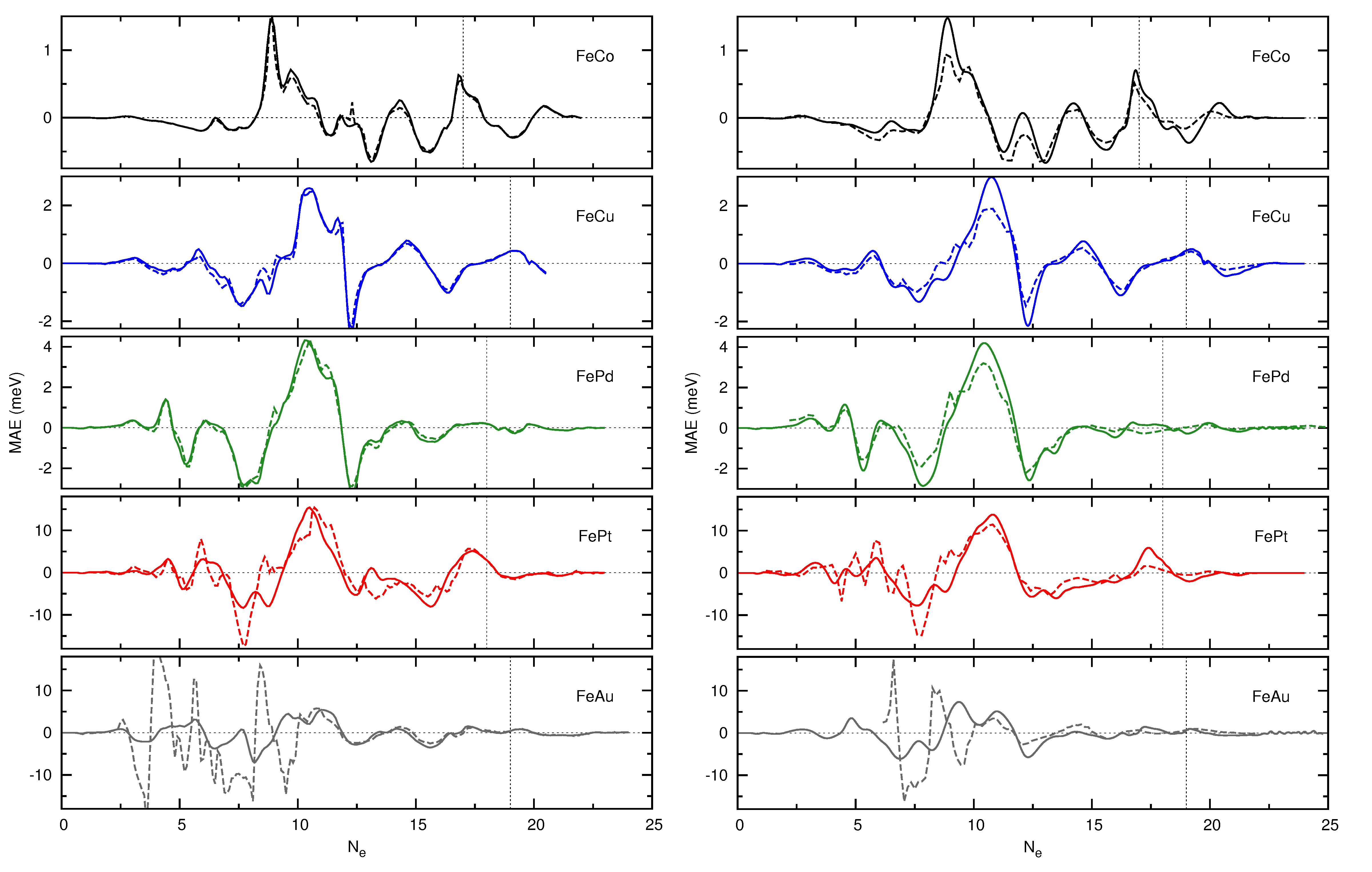}
\caption{\label{fig:kelly_vdl}
MAE as a function of the number of valence electrons $N_e$.
The vertical line indicates the position of the Fermi level in the
neutral systems. Solid (dashed) lines correspond to the NSCF (2PT).
Left: SG calculations, i.e. the dashed line is obtained from
Eq.~\ref{eq:vdllcao}, with a Fermi-Dirac smearing of $kT=10$\,meV
is used here.
Right:  VASP calculations, i.e. the dashed line is obtained from
Eq.~\ref{eq:vdl}, with a Fermi-Dirac smearing of $kT=50$\,meV.
}
\end{figure*}

Switching of the easy magnetization axis occurs several times as
a function of band filling in all cases. Interestingly, a MAE
reversal already takes place by removal or addition of one or two
electrons per unit cell.
Furthermore,
the MAE undergoes drastic changes in magnitude, even attaining
values which are one order of magnitude larger than the neutral ones,
specially in the $N_e\approx 10$ region where the $d$-bands show the
highest density of states.

With a methodological aim in mind, the study of the MAE in the
non-neutral case allows us to study the validity the 2PT perturbative
approach with greater confidence than in the neutral case,
since now we can compare a MAE curve with a rich 
structure instead of just a single value.
The SG 2PT curves (dashed lines in the left-hand panel of Fig.~\ref{fig:kelly_vdl})
are in very good agreement with
the NSCF curves for FeCo, FeCu, and FePd, while large discrepancies
appear when heavier atoms are present. 
This is particularly evident
in FeAu for $N_e=5-10$, which corresponds to the spectrum region
where the $d$-bands of Au lie.
Considering the strong dependence
of the MAE on the electron band structure details discussed in the
previous section and the fact that the agreement with NSCF is not
homogeneous as $N_e$ changes, both facts suggest that 2PT loses its validity
for elements with strong SOI. Thus, the coincidence in the
neutral-case FePt and FeAu MAEs could be considered 
to be fortuitous, in the sense that the coincidence is a consequence of 
the specific band structure of the alloy,
as it is the case of FePt (also observed in Ref.~\cite{bib:kota14}) and FeAu.
\footnote{In the FeAu alloy, the Au-$d$ states are mostly confined in a band 
between $-7$ and $-4$\,eV below the Fermi energy (see Supplementary 
Information Fig.~5). On the one hand, those states are 
subject to strong couplings by SOI, since $\xi_\mathrm{Au}=615$\,meV.
On the other hand, because of the $(\epsilon_{kn\sigma}-\epsilon_{kn'\sigma'})^{-1}$ factors,
those states have a weaker effect on the MAE for $N_e$ values close to 
charge neutrality ($N_e=19$) than for smaller $N_e$ values.
E.g. a band filling $N_e=10$ corresponds to a downward shift of the Fermi 
level of $-3.7$\,eV, close to the Au-$d$ states. 
Therefore, fast sharp oscillations are observed in the MAE curve at $N_e=5-10$,
while smooth behavior and apparent agreement with NSCF exists at $N_e>12$.
Importantly, this does not mean that the Au-$d$ states have a negligible
contribution, as evidenced by the absence of a plateau in the occupied states curve of 
the FeAu panel of Fig.~\ref{fig:fermiwinsiesta}, which represents the $N_e=19$ case.
For the FePt alloy, since the Pt-$d$ band is less localized in 
energies, the disagreement between 2PT and NSCF is visible throughout 
the $\mathrm{MAE}(N_e)$ curve.}

This restriction of the 2PT validity to light atoms is not a 
serious disadvantage. This approach
facilitates the MAE evaluation with fine $k$-point
grids and narrow smearing widths at a lower computational cost,
since it requires a single DFT calculation {\it without} SOI. 
We recall that a weak MCA does not necessarily follow from a small
$\xi$, as we see in the systems under study. In extended systems like
the current ones, band dispersion governs the MCA.

Both NSCF and 2PT describe SOI with a perturbative treatment of 
either the charge density changes or the $\xi$ parameter, respectively.
At this point, it is important to note another fundamental difference between
the NSCF and 2PT formalisms. In the former method, the eigenvalues
change with the magnetization axis and some degeneracies may be broken.
In the latter method, the unperturbed band structure
is not explicitely modified. Instead, $e-h$ pair excitations of the 
GS $| \Psi^{(0)} \rangle$ are induced by $H_{SO}$, with probabilities
given by their matrix elements. In other words, 2PT is a many-body
approach, while NSCF is a one-electron approach.
Thus, if we take the limit of single ions and uniaxial anisotropy, 
the 2PT approach described here tends to the 
widely-used formalism of the spin hamiltonian for non-degenerate states  
$H_{ion}=DS_z^2+E(S_x^2-S_y^2)$ \ \cite{bib:pryce50,bib:dai08}.

Visual evidence of the inherent difference between NSCF and 2PT 
is presented in Fig.~\ref{fig:feco_kresolved}. 
This figure shows, for the case of FeCo and the VASP-Wannier calculation, 
the MAE densities as a function of $N_e$ in the reciprocal space 
along a few high-symmetry directions inside the 1BZ.
To guide the eye, the bands without SOI have been transformed 
from energies to the corresponding filling $N_e$ and 
superimposed on the MAE density graph. As expected, for the NSCF case
(top panel) the regions of non-zero MAE are localized close to the unperturbed
bands, and take positive or negative values depending on the relative
$H_{SO}$ induced shifts in the eigenvalues between the $OZ$ and $OX$
magnetization directions. The map also reveals abrupt switching of the MAE
densities close to several band-crossings.
For example, this happens near the $\Gamma$ point and $N_e \simeq 9$, 
where a pair of majority spin bands is split by a few meV 
for magnetization along $OZ$.
Since the splitting is nearly symmetric in energy about 
the non-perturbed bands, the contribution to MAE is negative as 
the bottom band is  filled and changes sign when the
upper one starts to become occupied.
When both bands are completely filled the MAE vanishes.
In the 2PT approach the MAE density (lower panel) takes a very 
different aspect since the bandstructure is not modified
and, as shown by Fig.~\ref{fig:fermiwinsiesta}, 
$e-h$ pairs with an energy difference as large as a few eV
can contribute to MAE at a given $N_e$ (see also Supplementary Information Fig.~4).
Thus, the map presents
broad {plateaus with a non-negligible MAE density in areas devoid of bands, 
while sign changes are always
localized precisely at the bands since they take place
when the combined $e-h$ distribution functions (term 
$f(\epsilon_{kn\sigma})[1-f(\epsilon_{kn'\sigma'})]$ in Eq.~\ref{eq:de2kgrid})
also changes sign as the band is crossed.
Nevertheless, for FeCo the two approaches yield a similar overall
description of the MAE($N_e$) behaviour, 
as seen in the $k$-integrated curve of Fig.~\ref{fig:kelly_vdl} (right),
in spite of treating bandstructures in a different way by construction.

\begin{figure}
\includegraphics[width=\columnwidth]{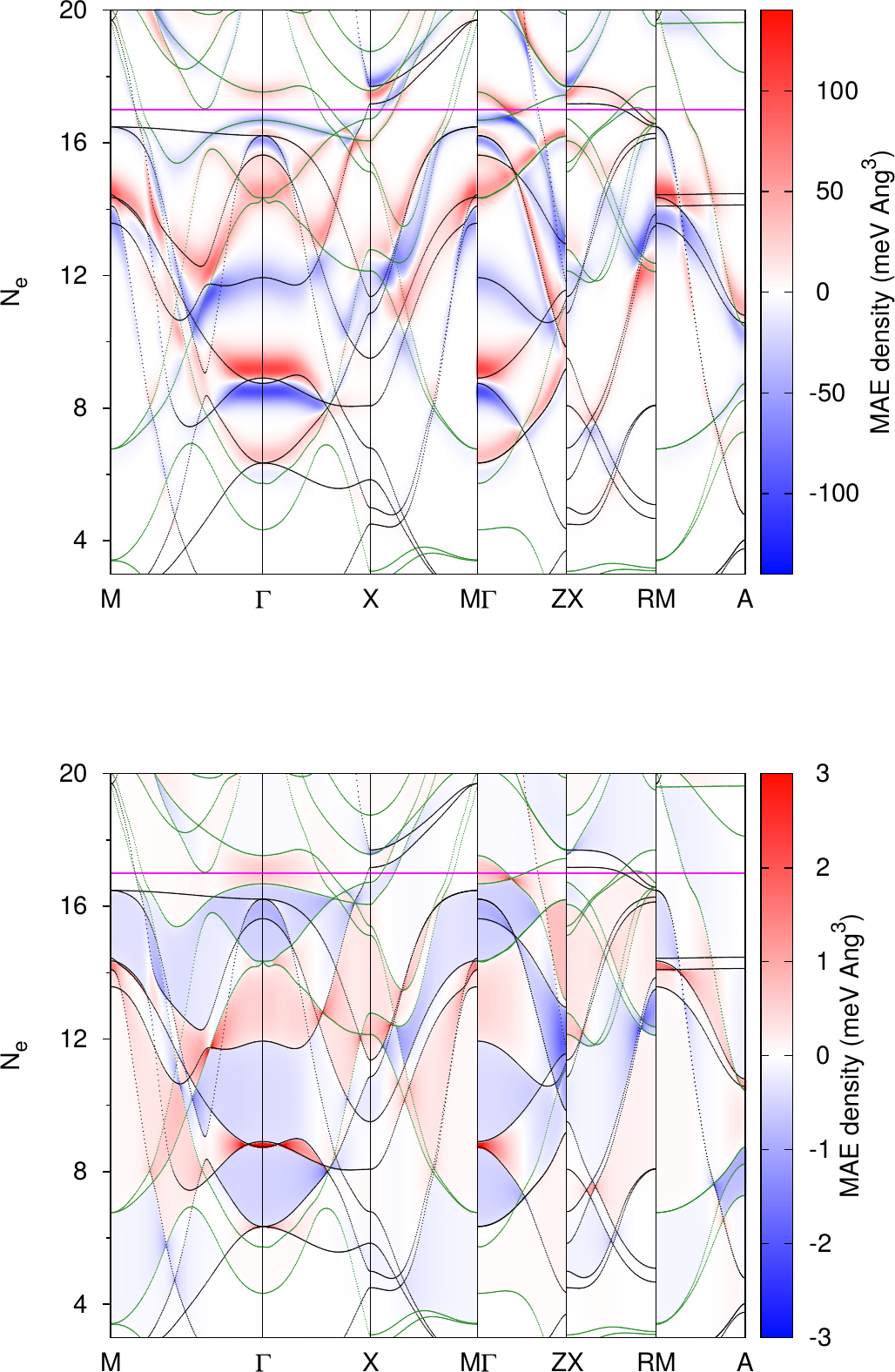}
\caption{\label{fig:feco_kresolved}
$k$-resolved MAE as a function of the number of valence electrons $N_e$
along high-symmetry directions inside the first Brillouin zone for
the FeCo alloy.
The top and bottom panels show the NSCF and 2PT approaches, respectively,
obtained from Eq.~\ref{eq:vdl} (VASP calculation)
with a Fermi-Dirac smearing of $kT=50$\,meV.
The correspondence between the Wannier-interpolated energy eigenvalues
without SOI and the number of valence electrons is shown as black
(majority spin) and green (minority spin) bandstructures.
The horizontal line indicates charge neutrality.
In the color scale, red and blue regions of the spectrum
account for a contribution to anisotropy easy axis along OZ or OX, respectively.
}
\end{figure}

With the 2PT calculation that uses MLWFs poorer agreement is found
in the profile of the $\mathrm{MAE}(N_e)$ curves, due to the limitations
of this methodology, still the qualitative behaviour is reproduced 
in all alloys except FeAu, where similar deviations as for the SG case
are found
(see right-hand panel of Fig.~\ref{fig:kelly_vdl}).
Therefore, it could be used under suitable conditions to make
predictions on the MAE behaviour as a function of doping or bias voltage
at a low computational cost from a DFT calculation which does not include SOI.
Those conditions are (i) weak SOI strength and (ii) resemblance between 
the MLWFs and $Y_{2m}$ spherical harmonics.
When the first condition is met, the MAE obtained in the on-site 
approximation (Eq.~\ref{eq:onsite}) performs as accurately as NSCF.
This has been checked with SG calculations (see Supplementary Information Fig.~7), 
where the drawback of point (ii) is not present.
Interestingly, the effect of inter-atomic $d$-orbital hybridization on the MAE is  
well captured by this methodology via the $F$ factors in
Eq.~\ref{eq:vdlcoefs}, while the crude approximation done for
the SOI matrix elements seems less important, 
as it can be deduced from the good agreement with the NSCF curves 
in the FeCo and FeCu panels of Fig.~\ref{fig:kelly_vdl} (right).

\section{Conclusions}
\label{sec:conclusions}

In this work, we have included the spin-orbit interaction (SOI) 
in DFT calculations at different levels of approximation to 
obtain the magnetocrystalline anisotropy energies (MAE).
As reference, we use fully-relativistic fully-self-consistent (SCF) 
calculations. We have examined the accuracy of 
(i) the so-called force theorem method, where the SOI 
is applied once to the charge density of a scalar-relativistic 
calculation, without subsequent self-consistency cycles (NSCF), and
(ii) a many-body second-order perturbation (2PT) treatment of the 
SOI on the scalar-relativistic ground state. 
With a compuational aim in mind, we have comfirmed that 
an accurate Fermi level determination is crucial 
to obtain converged MAE values.

As case studies, we have taken several 
FeMe tetragonal ferromagnetic alloys with $L1_0$ structure.
By choosing Me=Co, Cu, Pd, Pt, and Au, we cover the scenarios of
$s-d$ and $d-d$ hybrid band effects and a range of atomic 
SOI strength parameters $\xi$.
We find NSCF to provide reliable MAEs in general, whereas 
the 2PT approximation describes accurately the MAEs 
of FeCo, FeCu, ($\xi \sim 0.05$\,eV) and satisfactorily 
that of FePd ($\xi \sim 0.1$\,eV), but fails for the alloys 
containing $5d$ metals ($\xi \sim 0.5$\,eV). 
The difference in the performance of the two approximations 
has the following origin: NSCF is perturbative 
in the charge density changes by SOI, while 2PT is 
perturbative in $\xi$.

We find that the MAE in this family of alloys is 
determined not only by the empty spin-minority Fe states, 
which lie about 2\,eV above the Fermi level, but also by the 
whole valence band, which lies several eV below the Fermi level.
Thus, the magnetocrystalline anisotropy is
determined by electronic states that lie from the
Fermi level much further than the SOI strength parameter.
The details of the bandstructure are, in essence, 
responsible for the final MAE value.

Finally, the 2PT approximation is sound enough to
account for the anisotropy behaviour of the alloys under 
deviations from charge neutrality. We show that 
magnetic switching can be induced by addition or removal 
of electrons. This effect could be tuned by, for example, 
doping or strain, to find an scenario 
under a minimal bias voltage. These properties have ample 
applications in magnetoelectric and magnetostrictive devices.
The 2PT approximation to the SOI, when valid, can be used to 
study large numbers of these cases efficiently, as it uses 
as sole input a scalar-relativistic DFT calculation.

\begin{acknowledgments}
Discussions with G. Teobaldi  and M. dos Santos Dias are acknowledged.
M.B.-R. and A.A. thank financial support from
MINECO (grant number FIS2016-75862-P),
the University of the Basque Country (UPV/EHU) and 
the Basque Government (IT-756-13).
J.I.C. thanks MINECO for grant MAT2015-66888-C3-1R.
Computational resources were provided by the DIPC computing center.
\end{acknowledgments}

\bibliography{blanco}

\begin{thebibliography}{56}%
\makeatletter
\providecommand \@ifxundefined [1]{%
 \@ifx{#1\undefined}
}%
\providecommand \@ifnum [1]{%
 \ifnum #1\expandafter \@firstoftwo
 \else \expandafter \@secondoftwo
 \fi
}%
\providecommand \@ifx [1]{%
 \ifx #1\expandafter \@firstoftwo
 \else \expandafter \@secondoftwo
 \fi
}%
\providecommand \natexlab [1]{#1}%
\providecommand \enquote  [1]{``#1''}%
\providecommand \bibnamefont  [1]{#1}%
\providecommand \bibfnamefont [1]{#1}%
\providecommand \citenamefont [1]{#1}%
\providecommand \href@noop [0]{\@secondoftwo}%
\providecommand \href [0]{\begingroup \@sanitize@url \@href}%
\providecommand \@href[1]{\@@startlink{#1}\@@href}%
\providecommand \@@href[1]{\endgroup#1\@@endlink}%
\providecommand \@sanitize@url [0]{\catcode `\\12\catcode `\$12\catcode
  `\&12\catcode `\#12\catcode `\^12\catcode `\_12\catcode `\%12\relax}%
\providecommand \@@startlink[1]{}%
\providecommand \@@endlink[0]{}%
\providecommand \url  [0]{\begingroup\@sanitize@url \@url }%
\providecommand \@url [1]{\endgroup\@href {#1}{\urlprefix }}%
\providecommand \urlprefix  [0]{URL }%
\providecommand \Eprint [0]{\href }%
\providecommand \doibase [0]{http://dx.doi.org/}%
\providecommand \selectlanguage [0]{\@gobble}%
\providecommand \bibinfo  [0]{\@secondoftwo}%
\providecommand \bibfield  [0]{\@secondoftwo}%
\providecommand \translation [1]{[#1]}%
\providecommand \BibitemOpen [0]{}%
\providecommand \bibitemStop [0]{}%
\providecommand \bibitemNoStop [0]{.\EOS\space}%
\providecommand \EOS [0]{\spacefactor3000\relax}%
\providecommand \BibitemShut  [1]{\csname bibitem#1\endcsname}%
\let\auto@bib@innerbib\@empty
\bibitem [{\citenamefont {Moriya}(1960)}]{bib:moriya60}%
  \BibitemOpen
  \bibfield  {author} {\bibinfo {author} {\bibfnamefont {T.}~\bibnamefont
  {Moriya}},\ }\href {https://link.aps.org/doi/10.1103/PhysRev.120.91}
  {\bibfield  {journal} {\bibinfo  {journal} {Phys. Rev.}\ }\textbf {\bibinfo
  {volume} {120}},\ \bibinfo {pages} {91} (\bibinfo {year} {1960})}\BibitemShut
  {NoStop}%
\bibitem [{\citenamefont {Wang}\ \emph {et~al.}(2012)\citenamefont {Wang},
  \citenamefont {Li}, \citenamefont {Hageman},\ and\ \citenamefont
  {Chien}}]{bib:wang12}%
  \BibitemOpen
  \bibfield  {author} {\bibinfo {author} {\bibfnamefont {W.-G.}\ \bibnamefont
  {Wang}}, \bibinfo {author} {\bibfnamefont {M.}~\bibnamefont {Li}}, \bibinfo
  {author} {\bibfnamefont {S.}~\bibnamefont {Hageman}}, \ and\ \bibinfo
  {author} {\bibfnamefont {C.~L.}\ \bibnamefont {Chien}},\ }\href
  {http://dx.doi.org/10.1038/nmat3171} {\bibfield  {journal} {\bibinfo
  {journal} {Nature Materials}\ }\textbf {\bibinfo {volume} {11}},\ \bibinfo
  {pages} {64} (\bibinfo {year} {2012})}\BibitemShut {NoStop}%
\bibitem [{\citenamefont {Pantel}\ \emph {et~al.}(2012)\citenamefont {Pantel},
  \citenamefont {Goetze}, \citenamefont {Hesse},\ and\ \citenamefont
  {Alexe}}]{bib:pantel12}%
  \BibitemOpen
  \bibfield  {author} {\bibinfo {author} {\bibfnamefont {D.}~\bibnamefont
  {Pantel}}, \bibinfo {author} {\bibfnamefont {S.}~\bibnamefont {Goetze}},
  \bibinfo {author} {\bibfnamefont {D.}~\bibnamefont {Hesse}}, \ and\ \bibinfo
  {author} {\bibfnamefont {M.}~\bibnamefont {Alexe}},\ }\href
  {http://dx.doi.org/10.1038/nmat3254} {\bibfield  {journal} {\bibinfo
  {journal} {Nature Materials}\ }\textbf {\bibinfo {volume} {11}},\ \bibinfo
  {pages} {28} (\bibinfo {year} {2012})}\BibitemShut {NoStop}%
\bibitem [{\citenamefont {Dieny}\ and\ \citenamefont
  {Chshiev}(2017)}]{bib:dieny17}%
  \BibitemOpen
  \bibfield  {author} {\bibinfo {author} {\bibfnamefont {B.}~\bibnamefont
  {Dieny}}\ and\ \bibinfo {author} {\bibfnamefont {M.}~\bibnamefont
  {Chshiev}},\ }\href {https://link.aps.org/doi/10.1103/RevModPhys.89.025008}
  {\bibfield  {journal} {\bibinfo  {journal} {Rev. Mod. Phys.}\ }\textbf
  {\bibinfo {volume} {89}},\ \bibinfo {pages} {025008} (\bibinfo {year}
  {2017})}\BibitemShut {NoStop}%
\bibitem [{\citenamefont {Mermin}\ and\ \citenamefont
  {Wagner}(1966)}]{bib:mermin66}%
  \BibitemOpen
  \bibfield  {author} {\bibinfo {author} {\bibfnamefont {N.~D.}\ \bibnamefont
  {Mermin}}\ and\ \bibinfo {author} {\bibfnamefont {H.}~\bibnamefont
  {Wagner}},\ }\href {https://link.aps.org/doi/10.1103/PhysRevLett.17.1133}
  {\bibfield  {journal} {\bibinfo  {journal} {Phys. Rev. Lett.}\ }\textbf
  {\bibinfo {volume} {17}},\ \bibinfo {pages} {1133} (\bibinfo {year}
  {1966})}\BibitemShut {NoStop}%
\bibitem [{\citenamefont {Jia}\ \emph {et~al.}(2012)\citenamefont {Jia},
  \citenamefont {Sukhov}, \citenamefont {Horley},\ and\ \citenamefont
  {Berakdar}}]{bib:jia12}%
  \BibitemOpen
  \bibfield  {author} {\bibinfo {author} {\bibfnamefont {C.}~\bibnamefont
  {Jia}}, \bibinfo {author} {\bibfnamefont {A.}~\bibnamefont {Sukhov}},
  \bibinfo {author} {\bibfnamefont {P.~P.}\ \bibnamefont {Horley}}, \ and\
  \bibinfo {author} {\bibfnamefont {J.}~\bibnamefont {Berakdar}},\ }\href
  {http://stacks.iop.org/0295-5075/99/i=1/a=17004} {\bibfield  {journal}
  {\bibinfo  {journal} {EPL (Europhysics Letters)}\ }\textbf {\bibinfo {volume}
  {99}},\ \bibinfo {pages} {17004} (\bibinfo {year} {2012})}\BibitemShut
  {NoStop}%
\bibitem [{\citenamefont {Duan}\ \emph
  {et~al.}(2008{\natexlab{a}})\citenamefont {Duan}, \citenamefont {Velev},
  \citenamefont {Sabirianov}, \citenamefont {Mei}, \citenamefont {Jaswal},\
  and\ \citenamefont {Tsymbal}}]{bib:duan08b}%
  \BibitemOpen
  \bibfield  {author} {\bibinfo {author} {\bibfnamefont {C.-G.}\ \bibnamefont
  {Duan}}, \bibinfo {author} {\bibfnamefont {J.~P.}\ \bibnamefont {Velev}},
  \bibinfo {author} {\bibfnamefont {R.~F.}\ \bibnamefont {Sabirianov}},
  \bibinfo {author} {\bibfnamefont {W.~N.}\ \bibnamefont {Mei}}, \bibinfo
  {author} {\bibfnamefont {S.~S.}\ \bibnamefont {Jaswal}}, \ and\ \bibinfo
  {author} {\bibfnamefont {E.~Y.}\ \bibnamefont {Tsymbal}},\ }\href
  {https://doi.org/10.1063/1.2901879} {\bibfield  {journal} {\bibinfo
  {journal} {Applied Physics Letters}\ }\textbf {\bibinfo {volume} {92}},\
  \bibinfo {pages} {122905} (\bibinfo {year} {2008}{\natexlab{a}})}\BibitemShut
  {NoStop}%
\bibitem [{\citenamefont {Faleev}\ \emph {et~al.}(2017)\citenamefont {Faleev},
  \citenamefont {Ferrante}, \citenamefont {Jeong}, \citenamefont {Samant},
  \citenamefont {Jones},\ and\ \citenamefont {Parkin}}]{bib:faleev17}%
  \BibitemOpen
  \bibfield  {author} {\bibinfo {author} {\bibfnamefont {S.~V.}\ \bibnamefont
  {Faleev}}, \bibinfo {author} {\bibfnamefont {Y.}~\bibnamefont {Ferrante}},
  \bibinfo {author} {\bibfnamefont {J.}~\bibnamefont {Jeong}}, \bibinfo
  {author} {\bibfnamefont {M.~G.}\ \bibnamefont {Samant}}, \bibinfo {author}
  {\bibfnamefont {B.}~\bibnamefont {Jones}}, \ and\ \bibinfo {author}
  {\bibfnamefont {S.~S.~P.}\ \bibnamefont {Parkin}},\ }\href
  {https://link.aps.org/doi/10.1103/PhysRevMaterials.1.024402} {\bibfield
  {journal} {\bibinfo  {journal} {Phys. Rev. Materials}\ }\textbf {\bibinfo
  {volume} {1}},\ \bibinfo {pages} {024402} (\bibinfo {year}
  {2017})}\BibitemShut {NoStop}%
\bibitem [{\citenamefont {Herper}(2018)}]{bib:herper18}%
  \BibitemOpen
  \bibfield  {author} {\bibinfo {author} {\bibfnamefont {H.~C.}\ \bibnamefont
  {Herper}},\ }\href {https://link.aps.org/doi/10.1103/PhysRevB.98.014411}
  {\bibfield  {journal} {\bibinfo  {journal} {Phys. Rev. B}\ }\textbf {\bibinfo
  {volume} {98}},\ \bibinfo {pages} {014411} (\bibinfo {year}
  {2018})}\BibitemShut {NoStop}%
\bibitem [{\citenamefont {Zhang}\ \emph {et~al.}(2010)\citenamefont {Zhang},
  \citenamefont {Cao},\ and\ \citenamefont {Wu}}]{bib:zhang10}%
  \BibitemOpen
  \bibfield  {author} {\bibinfo {author} {\bibfnamefont {Y.~N.}\ \bibnamefont
  {Zhang}}, \bibinfo {author} {\bibfnamefont {J.~X.}\ \bibnamefont {Cao}}, \
  and\ \bibinfo {author} {\bibfnamefont {R.~Q.}\ \bibnamefont {Wu}},\ }\href
  {http://dx.doi.org/10.1063/1.3318420} {\bibfield  {journal} {\bibinfo
  {journal} {Appl. Phys. Lett.}\ }\textbf {\bibinfo {volume} {96}},\ \bibinfo
  {pages} {062508} (\bibinfo {year} {2010})}\BibitemShut {NoStop}%
\bibitem [{\citenamefont {Wang}\ \emph {et~al.}(2013)\citenamefont {Wang},
  \citenamefont {Zhang}, \citenamefont {Wu}, \citenamefont {Sun}, \citenamefont
  {Xu},\ and\ \citenamefont {Zhang}}]{bib:wang13}%
  \BibitemOpen
  \bibfield  {author} {\bibinfo {author} {\bibfnamefont {H.}~\bibnamefont
  {Wang}}, \bibinfo {author} {\bibfnamefont {Y.~N.}\ \bibnamefont {Zhang}},
  \bibinfo {author} {\bibfnamefont {R.~Q.}\ \bibnamefont {Wu}}, \bibinfo
  {author} {\bibfnamefont {L.~Z.}\ \bibnamefont {Sun}}, \bibinfo {author}
  {\bibfnamefont {D.~S.}\ \bibnamefont {Xu}}, \ and\ \bibinfo {author}
  {\bibfnamefont {Z.~D.}\ \bibnamefont {Zhang}},\ }\href
  {http://dx.doi.org/10.1038/srep03521} {\bibfield  {journal} {\bibinfo
  {journal} {Scientific Reports}\ }\textbf {\bibinfo {volume} {3}},\ \bibinfo
  {pages} {3521} (\bibinfo {year} {2013})}\BibitemShut {NoStop}%
\bibitem [{\citenamefont {Burkert}\ \emph {et~al.}(2004)\citenamefont
  {Burkert}, \citenamefont {Nordstr\"om}, \citenamefont {Eriksson},\ and\
  \citenamefont {Heinonen}}]{bib:burkert04}%
  \BibitemOpen
  \bibfield  {author} {\bibinfo {author} {\bibfnamefont {T.}~\bibnamefont
  {Burkert}}, \bibinfo {author} {\bibfnamefont {L.}~\bibnamefont
  {Nordstr\"om}}, \bibinfo {author} {\bibfnamefont {O.}~\bibnamefont
  {Eriksson}}, \ and\ \bibinfo {author} {\bibfnamefont {O.}~\bibnamefont
  {Heinonen}},\ }\href {https://link.aps.org/doi/10.1103/PhysRevLett.93.027203}
  {\bibfield  {journal} {\bibinfo  {journal} {Phys. Rev. Lett.}\ }\textbf
  {\bibinfo {volume} {93}},\ \bibinfo {pages} {027203} (\bibinfo {year}
  {2004})}\BibitemShut {NoStop}%
\bibitem [{\citenamefont {Andersson}\ \emph {et~al.}(2006)\citenamefont
  {Andersson}, \citenamefont {Burkert}, \citenamefont {Warnicke}, \citenamefont
  {Bj\"orck}, \citenamefont {Sanyal}, \citenamefont {Chacon}, \citenamefont
  {Zlotea}, \citenamefont {Nordstr\"om}, \citenamefont {Nordblad},\ and\
  \citenamefont {Eriksson}}]{bib:andersson06}%
  \BibitemOpen
  \bibfield  {author} {\bibinfo {author} {\bibfnamefont {G.}~\bibnamefont
  {Andersson}}, \bibinfo {author} {\bibfnamefont {T.}~\bibnamefont {Burkert}},
  \bibinfo {author} {\bibfnamefont {P.}~\bibnamefont {Warnicke}}, \bibinfo
  {author} {\bibfnamefont {M.}~\bibnamefont {Bj\"orck}}, \bibinfo {author}
  {\bibfnamefont {B.}~\bibnamefont {Sanyal}}, \bibinfo {author} {\bibfnamefont
  {C.}~\bibnamefont {Chacon}}, \bibinfo {author} {\bibfnamefont
  {C.}~\bibnamefont {Zlotea}}, \bibinfo {author} {\bibfnamefont
  {L.}~\bibnamefont {Nordstr\"om}}, \bibinfo {author} {\bibfnamefont
  {P.}~\bibnamefont {Nordblad}}, \ and\ \bibinfo {author} {\bibfnamefont
  {O.}~\bibnamefont {Eriksson}},\ }\href
  {https://link.aps.org/doi/10.1103/PhysRevLett.96.037205} {\bibfield
  {journal} {\bibinfo  {journal} {Phys. Rev. Lett.}\ }\textbf {\bibinfo
  {volume} {96}},\ \bibinfo {pages} {037205} (\bibinfo {year}
  {2006})}\BibitemShut {NoStop}%
\bibitem [{\citenamefont {Turek}\ \emph {et~al.}(2012)\citenamefont {Turek},
  \citenamefont {Kudrnovsk\'y},\ and\ \citenamefont {Carva}}]{bib:turek12}%
  \BibitemOpen
  \bibfield  {author} {\bibinfo {author} {\bibfnamefont {I.}~\bibnamefont
  {Turek}}, \bibinfo {author} {\bibfnamefont {J.}~\bibnamefont {Kudrnovsk\'y}},
  \ and\ \bibinfo {author} {\bibfnamefont {K.}~\bibnamefont {Carva}},\ }\href
  {https://link.aps.org/doi/10.1103/PhysRevB.86.174430} {\bibfield  {journal}
  {\bibinfo  {journal} {Phys. Rev. B}\ }\textbf {\bibinfo {volume} {86}},\
  \bibinfo {pages} {174430} (\bibinfo {year} {2012})}\BibitemShut {NoStop}%
\bibitem [{\citenamefont {Quintana}\ \emph {et~al.}(2017)\citenamefont
  {Quintana}, \citenamefont {Zhang}, \citenamefont {Isarain-Ch\'avez},
  \citenamefont {Men\'endez}, \citenamefont {Cuadrado}, \citenamefont {Robles},
  \citenamefont {Bar\'o}, \citenamefont {Guerrero}, \citenamefont {Pan\'e},
  \citenamefont {Nelson}, \citenamefont {M\'uller}, \citenamefont {Ordej\'on},
  \citenamefont {Nogu\'es}, \citenamefont {Pellicer},\ and\ \citenamefont
  {Sort}}]{bib:quintana17}%
  \BibitemOpen
  \bibfield  {author} {\bibinfo {author} {\bibfnamefont {A.}~\bibnamefont
  {Quintana}}, \bibinfo {author} {\bibfnamefont {J.}~\bibnamefont {Zhang}},
  \bibinfo {author} {\bibfnamefont {E.}~\bibnamefont {Isarain-Ch\'avez}},
  \bibinfo {author} {\bibfnamefont {E.}~\bibnamefont {Men\'endez}}, \bibinfo
  {author} {\bibfnamefont {R.}~\bibnamefont {Cuadrado}}, \bibinfo {author}
  {\bibfnamefont {R.}~\bibnamefont {Robles}}, \bibinfo {author} {\bibfnamefont
  {M.~D.}\ \bibnamefont {Bar\'o}}, \bibinfo {author} {\bibfnamefont
  {M.}~\bibnamefont {Guerrero}}, \bibinfo {author} {\bibfnamefont
  {S.}~\bibnamefont {Pan\'e}}, \bibinfo {author} {\bibfnamefont {B.~J.}\
  \bibnamefont {Nelson}}, \bibinfo {author} {\bibfnamefont {C.~M.}\
  \bibnamefont {M\'uller}}, \bibinfo {author} {\bibfnamefont {P.}~\bibnamefont
  {Ordej\'on}}, \bibinfo {author} {\bibfnamefont {J.}~\bibnamefont {Nogu\'es}},
  \bibinfo {author} {\bibfnamefont {E.}~\bibnamefont {Pellicer}}, \ and\
  \bibinfo {author} {\bibfnamefont {J.}~\bibnamefont {Sort}},\ }\href
  {https://onlinelibrary.wiley.com/doi/abs/10.1002/adfm.201701904} {\bibfield
  {journal} {\bibinfo  {journal} {Advanced Functional Materials}\ }\textbf
  {\bibinfo {volume} {27}},\ \bibinfo {pages} {1701904} (\bibinfo {year}
  {2017})}\BibitemShut {NoStop}%
\bibitem [{\citenamefont {Duan}\ \emph
  {et~al.}(2008{\natexlab{b}})\citenamefont {Duan}, \citenamefont {Velev},
  \citenamefont {Sabirianov}, \citenamefont {Zhu}, \citenamefont {Chu},
  \citenamefont {Jaswal},\ and\ \citenamefont {Tsymbal}}]{bib:duan08}%
  \BibitemOpen
  \bibfield  {author} {\bibinfo {author} {\bibfnamefont {C.-G.}\ \bibnamefont
  {Duan}}, \bibinfo {author} {\bibfnamefont {J.~P.}\ \bibnamefont {Velev}},
  \bibinfo {author} {\bibfnamefont {R.~F.}\ \bibnamefont {Sabirianov}},
  \bibinfo {author} {\bibfnamefont {Z.}~\bibnamefont {Zhu}}, \bibinfo {author}
  {\bibfnamefont {J.}~\bibnamefont {Chu}}, \bibinfo {author} {\bibfnamefont
  {S.~S.}\ \bibnamefont {Jaswal}}, \ and\ \bibinfo {author} {\bibfnamefont
  {E.~Y.}\ \bibnamefont {Tsymbal}},\ }\href
  {https://link.aps.org/doi/10.1103/PhysRevLett.101.137201} {\bibfield
  {journal} {\bibinfo  {journal} {Phys. Rev. Lett.}\ }\textbf {\bibinfo
  {volume} {101}},\ \bibinfo {pages} {137201} (\bibinfo {year}
  {2008}{\natexlab{b}})}\BibitemShut {NoStop}%
\bibitem [{\citenamefont {Nakamura}\ \emph {et~al.}(2009)\citenamefont
  {Nakamura}, \citenamefont {Shimabukuro}, \citenamefont {Fujiwara},
  \citenamefont {Akiyama}, \citenamefont {Ito},\ and\ \citenamefont
  {Freeman}}]{bib:nakamura09}%
  \BibitemOpen
  \bibfield  {author} {\bibinfo {author} {\bibfnamefont {K.}~\bibnamefont
  {Nakamura}}, \bibinfo {author} {\bibfnamefont {R.}~\bibnamefont
  {Shimabukuro}}, \bibinfo {author} {\bibfnamefont {Y.}~\bibnamefont
  {Fujiwara}}, \bibinfo {author} {\bibfnamefont {T.}~\bibnamefont {Akiyama}},
  \bibinfo {author} {\bibfnamefont {T.}~\bibnamefont {Ito}}, \ and\ \bibinfo
  {author} {\bibfnamefont {A.~J.}\ \bibnamefont {Freeman}},\ }\href
  {https://link.aps.org/doi/10.1103/PhysRevLett.102.187201} {\bibfield
  {journal} {\bibinfo  {journal} {Phys. Rev. Lett.}\ }\textbf {\bibinfo
  {volume} {102}},\ \bibinfo {pages} {187201} (\bibinfo {year}
  {2009})}\BibitemShut {NoStop}%
\bibitem [{\citenamefont {Gamble}\ \emph {et~al.}(2009)\citenamefont {Gamble},
  \citenamefont {Burkhardt}, \citenamefont {Kashuba}, \citenamefont
  {Allenspach}, \citenamefont {Parkin}, \citenamefont {Siegmann},\ and\
  \citenamefont {St\"ohr}}]{bib:gamble09}%
  \BibitemOpen
  \bibfield  {author} {\bibinfo {author} {\bibfnamefont {S.~J.}\ \bibnamefont
  {Gamble}}, \bibinfo {author} {\bibfnamefont {M.~H.}\ \bibnamefont
  {Burkhardt}}, \bibinfo {author} {\bibfnamefont {A.}~\bibnamefont {Kashuba}},
  \bibinfo {author} {\bibfnamefont {R.}~\bibnamefont {Allenspach}}, \bibinfo
  {author} {\bibfnamefont {S.~S.~P.}\ \bibnamefont {Parkin}}, \bibinfo {author}
  {\bibfnamefont {H.~C.}\ \bibnamefont {Siegmann}}, \ and\ \bibinfo {author}
  {\bibfnamefont {J.}~\bibnamefont {St\"ohr}},\ }\href
  {https://link.aps.org/doi/10.1103/PhysRevLett.102.217201} {\bibfield
  {journal} {\bibinfo  {journal} {Phys. Rev. Lett.}\ }\textbf {\bibinfo
  {volume} {102}},\ \bibinfo {pages} {217201} (\bibinfo {year}
  {2009})}\BibitemShut {NoStop}%
\bibitem [{\citenamefont {Daalderop}\ \emph {et~al.}(1990)\citenamefont
  {Daalderop}, \citenamefont {Kelly},\ and\ \citenamefont
  {Schuurmans}}]{bib:daalderop90}%
  \BibitemOpen
  \bibfield  {author} {\bibinfo {author} {\bibfnamefont {G.~H.~O.}\
  \bibnamefont {Daalderop}}, \bibinfo {author} {\bibfnamefont {P.~J.}\
  \bibnamefont {Kelly}}, \ and\ \bibinfo {author} {\bibfnamefont {M.~F.~H.}\
  \bibnamefont {Schuurmans}},\ }\href
  {https://link.aps.org/doi/10.1103/PhysRevB.41.11919} {\bibfield  {journal}
  {\bibinfo  {journal} {Phys. Rev. B}\ }\textbf {\bibinfo {volume} {41}},\
  \bibinfo {pages} {11919} (\bibinfo {year} {1990})}\BibitemShut {NoStop}%
\bibitem [{\citenamefont {Li}\ \emph {et~al.}(1990)\citenamefont {Li},
  \citenamefont {Freeman}, \citenamefont {Jansen},\ and\ \citenamefont
  {Fu}}]{bib:li90}%
  \BibitemOpen
  \bibfield  {author} {\bibinfo {author} {\bibfnamefont {C.}~\bibnamefont
  {Li}}, \bibinfo {author} {\bibfnamefont {A.~J.}\ \bibnamefont {Freeman}},
  \bibinfo {author} {\bibfnamefont {H.~J.~F.}\ \bibnamefont {Jansen}}, \ and\
  \bibinfo {author} {\bibfnamefont {C.~L.}\ \bibnamefont {Fu}},\ }\href
  {https://link.aps.org/doi/10.1103/PhysRevB.42.5433} {\bibfield  {journal}
  {\bibinfo  {journal} {Phys. Rev. B}\ }\textbf {\bibinfo {volume} {42}},\
  \bibinfo {pages} {5433} (\bibinfo {year} {1990})}\BibitemShut {NoStop}%
\bibitem [{\citenamefont {Weinert}\ \emph {et~al.}(1985)\citenamefont
  {Weinert}, \citenamefont {Watson},\ and\ \citenamefont
  {Davenport}}]{bib:weinert85}%
  \BibitemOpen
  \bibfield  {author} {\bibinfo {author} {\bibfnamefont {M.}~\bibnamefont
  {Weinert}}, \bibinfo {author} {\bibfnamefont {R.~E.}\ \bibnamefont {Watson}},
  \ and\ \bibinfo {author} {\bibfnamefont {J.~W.}\ \bibnamefont {Davenport}},\
  }\href {https://link.aps.org/doi/10.1103/PhysRevB.32.2115} {\bibfield
  {journal} {\bibinfo  {journal} {Phys. Rev. B}\ }\textbf {\bibinfo {volume}
  {32}},\ \bibinfo {pages} {2115} (\bibinfo {year} {1985})}\BibitemShut
  {NoStop}%
\bibitem [{\citenamefont {Bruno}(1989)}]{bib:bruno89}%
  \BibitemOpen
  \bibfield  {author} {\bibinfo {author} {\bibfnamefont {P.}~\bibnamefont
  {Bruno}},\ }\href {https://link.aps.org/doi/10.1103/PhysRevB.39.865}
  {\bibfield  {journal} {\bibinfo  {journal} {Phys. Rev. B}\ }\textbf {\bibinfo
  {volume} {39}},\ \bibinfo {pages} {865} (\bibinfo {year} {1989})}\BibitemShut
  {NoStop}%
\bibitem [{\citenamefont {Cinal}\ \emph {et~al.}(1994)\citenamefont {Cinal},
  \citenamefont {Edwards},\ and\ \citenamefont {Mathon}}]{bib:cinal94}%
  \BibitemOpen
  \bibfield  {author} {\bibinfo {author} {\bibfnamefont {M.}~\bibnamefont
  {Cinal}}, \bibinfo {author} {\bibfnamefont {D.~M.}\ \bibnamefont {Edwards}},
  \ and\ \bibinfo {author} {\bibfnamefont {J.}~\bibnamefont {Mathon}},\ }\href
  {https://link.aps.org/doi/10.1103/PhysRevB.50.3754} {\bibfield  {journal}
  {\bibinfo  {journal} {Phys. Rev. B}\ }\textbf {\bibinfo {volume} {50}},\
  \bibinfo {pages} {3754} (\bibinfo {year} {1994})}\BibitemShut {NoStop}%
\bibitem [{\citenamefont {van~der Laan}(1998)}]{bib:vanderlaan98}%
  \BibitemOpen
  \bibfield  {author} {\bibinfo {author} {\bibfnamefont {G.}~\bibnamefont
  {van~der Laan}},\ }\href {http://stacks.iop.org/0953-8984/10/i=14/a=012}
  {\bibfield  {journal} {\bibinfo  {journal} {Journal of Physics: Condensed
  Matter}\ }\textbf {\bibinfo {volume} {10}},\ \bibinfo {pages} {3239}
  (\bibinfo {year} {1998})}\BibitemShut {NoStop}%
\bibitem [{\citenamefont {Takayama}\ \emph {et~al.}(1976)\citenamefont
  {Takayama}, \citenamefont {Bohnen},\ and\ \citenamefont
  {Fulde}}]{bib:takayama76}%
  \BibitemOpen
  \bibfield  {author} {\bibinfo {author} {\bibfnamefont {H.}~\bibnamefont
  {Takayama}}, \bibinfo {author} {\bibfnamefont {K.-P.}\ \bibnamefont
  {Bohnen}}, \ and\ \bibinfo {author} {\bibfnamefont {P.}~\bibnamefont
  {Fulde}},\ }\href {https://link.aps.org/doi/10.1103/PhysRevB.14.2287}
  {\bibfield  {journal} {\bibinfo  {journal} {Phys. Rev. B}\ }\textbf {\bibinfo
  {volume} {14}},\ \bibinfo {pages} {2287} (\bibinfo {year}
  {1976})}\BibitemShut {NoStop}%
\bibitem [{\citenamefont {Solovyev}\ \emph {et~al.}(1995)\citenamefont
  {Solovyev}, \citenamefont {Dederichs},\ and\ \citenamefont
  {Mertig}}]{bib:solovyev95}%
  \BibitemOpen
  \bibfield  {author} {\bibinfo {author} {\bibfnamefont {I.~V.}\ \bibnamefont
  {Solovyev}}, \bibinfo {author} {\bibfnamefont {P.~H.}\ \bibnamefont
  {Dederichs}}, \ and\ \bibinfo {author} {\bibfnamefont {I.}~\bibnamefont
  {Mertig}},\ }\href {https://link.aps.org/doi/10.1103/PhysRevB.52.13419}
  {\bibfield  {journal} {\bibinfo  {journal} {Phys. Rev. B}\ }\textbf {\bibinfo
  {volume} {52}},\ \bibinfo {pages} {13419} (\bibinfo {year}
  {1995})}\BibitemShut {NoStop}%
\bibitem [{\citenamefont {Ke}\ and\ \citenamefont {van
  Schilfgaarde}(2015)}]{bib:ke15}%
  \BibitemOpen
  \bibfield  {author} {\bibinfo {author} {\bibfnamefont {L.}~\bibnamefont
  {Ke}}\ and\ \bibinfo {author} {\bibfnamefont {M.}~\bibnamefont {van
  Schilfgaarde}},\ }\href {https://link.aps.org/doi/10.1103/PhysRevB.92.014423}
  {\bibfield  {journal} {\bibinfo  {journal} {Phys. Rev. B}\ }\textbf {\bibinfo
  {volume} {92}},\ \bibinfo {pages} {014423} (\bibinfo {year}
  {2015})}\BibitemShut {NoStop}%
\bibitem [{\citenamefont {Inoue}\ \emph {et~al.}(2015)\citenamefont {Inoue},
  \citenamefont {Yoshioka},\ and\ \citenamefont {Tsuchiura}}]{bib:inoue15}%
  \BibitemOpen
  \bibfield  {author} {\bibinfo {author} {\bibfnamefont {J.-i.}\ \bibnamefont
  {Inoue}}, \bibinfo {author} {\bibfnamefont {T.}~\bibnamefont {Yoshioka}}, \
  and\ \bibinfo {author} {\bibfnamefont {H.}~\bibnamefont {Tsuchiura}},\ }\href
  {https://doi.org/10.1063/1.4913898} {\bibfield  {journal} {\bibinfo
  {journal} {Journal of Applied Physics}\ }\textbf {\bibinfo {volume} {117}},\
  \bibinfo {pages} {17C720} (\bibinfo {year} {2015})}\BibitemShut {NoStop}%
\bibitem [{\citenamefont {Wang}\ \emph {et~al.}(1993)\citenamefont {Wang},
  \citenamefont {Wu},\ and\ \citenamefont {Freeman}}]{bib:wang93}%
  \BibitemOpen
  \bibfield  {author} {\bibinfo {author} {\bibfnamefont {D.-s.}\ \bibnamefont
  {Wang}}, \bibinfo {author} {\bibfnamefont {R.}~\bibnamefont {Wu}}, \ and\
  \bibinfo {author} {\bibfnamefont {A.~J.}\ \bibnamefont {Freeman}},\ }\href
  {https://link.aps.org/doi/10.1103/PhysRevLett.70.869} {\bibfield  {journal}
  {\bibinfo  {journal} {Phys. Rev. Lett.}\ }\textbf {\bibinfo {volume} {70}},\
  \bibinfo {pages} {869} (\bibinfo {year} {1993})}\BibitemShut {NoStop}%
\bibitem [{\citenamefont {Abate}\ and\ \citenamefont
  {Asdente}(1965)}]{bib:abate65}%
  \BibitemOpen
  \bibfield  {author} {\bibinfo {author} {\bibfnamefont {E.}~\bibnamefont
  {Abate}}\ and\ \bibinfo {author} {\bibfnamefont {M.}~\bibnamefont
  {Asdente}},\ }\href {https://link.aps.org/doi/10.1103/PhysRev.140.A1303}
  {\bibfield  {journal} {\bibinfo  {journal} {Phys. Rev.}\ }\textbf {\bibinfo
  {volume} {140}},\ \bibinfo {pages} {A1303} (\bibinfo {year}
  {1965})}\BibitemShut {NoStop}%
\bibitem [{\citenamefont {Els{\"a}sser}\ \emph {et~al.}(1988)\citenamefont
  {Els{\"a}sser}, \citenamefont {F{\"a}hnle}, \citenamefont {Brandt},\ and\
  \citenamefont {B{\"o}hm}}]{bib:elsasser88}%
  \BibitemOpen
  \bibfield  {author} {\bibinfo {author} {\bibfnamefont {C.}~\bibnamefont
  {Els{\"a}sser}}, \bibinfo {author} {\bibfnamefont {M.}~\bibnamefont
  {F{\"a}hnle}}, \bibinfo {author} {\bibfnamefont {E.~H.}\ \bibnamefont
  {Brandt}}, \ and\ \bibinfo {author} {\bibfnamefont {M.~C.}\ \bibnamefont
  {B{\"o}hm}},\ }\href {http://stacks.iop.org/0305-4608/18/i=11/a=018}
  {\bibfield  {journal} {\bibinfo  {journal} {Journal of Physics F: Metal
  Physics}\ }\textbf {\bibinfo {volume} {18}},\ \bibinfo {pages} {2463}
  (\bibinfo {year} {1988})}\BibitemShut {NoStop}%
\bibitem [{\citenamefont {Marzari}\ and\ \citenamefont
  {Vanderbilt}(1997)}]{bib:marzari97}%
  \BibitemOpen
  \bibfield  {author} {\bibinfo {author} {\bibfnamefont {N.}~\bibnamefont
  {Marzari}}\ and\ \bibinfo {author} {\bibfnamefont {D.}~\bibnamefont
  {Vanderbilt}},\ }\href {https://link.aps.org/doi/10.1103/PhysRevB.56.12847}
  {\bibfield  {journal} {\bibinfo  {journal} {Phys. Rev. B}\ }\textbf {\bibinfo
  {volume} {56}},\ \bibinfo {pages} {12847} (\bibinfo {year}
  {1997})}\BibitemShut {NoStop}%
\bibitem [{\citenamefont {Souza}\ \emph {et~al.}(2001)\citenamefont {Souza},
  \citenamefont {Marzari},\ and\ \citenamefont {Vanderbilt}}]{bib:souza01}%
  \BibitemOpen
  \bibfield  {author} {\bibinfo {author} {\bibfnamefont {I.}~\bibnamefont
  {Souza}}, \bibinfo {author} {\bibfnamefont {N.}~\bibnamefont {Marzari}}, \
  and\ \bibinfo {author} {\bibfnamefont {D.}~\bibnamefont {Vanderbilt}},\
  }\href {https://link.aps.org/doi/10.1103/PhysRevB.65.035109} {\bibfield
  {journal} {\bibinfo  {journal} {Phys. Rev. B}\ }\textbf {\bibinfo {volume}
  {65}},\ \bibinfo {pages} {035109} (\bibinfo {year} {2001})}\BibitemShut
  {NoStop}%
\bibitem [{\citenamefont {Yates}\ \emph {et~al.}(2007)\citenamefont {Yates},
  \citenamefont {Wang}, \citenamefont {Vanderbilt},\ and\ \citenamefont
  {Souza}}]{bib:yates07}%
  \BibitemOpen
  \bibfield  {author} {\bibinfo {author} {\bibfnamefont {J.~R.}\ \bibnamefont
  {Yates}}, \bibinfo {author} {\bibfnamefont {X.}~\bibnamefont {Wang}},
  \bibinfo {author} {\bibfnamefont {D.}~\bibnamefont {Vanderbilt}}, \ and\
  \bibinfo {author} {\bibfnamefont {I.}~\bibnamefont {Souza}},\ }\href
  {https://link.aps.org/doi/10.1103/PhysRevB.75.195121} {\bibfield  {journal}
  {\bibinfo  {journal} {Phys. Rev. B}\ }\textbf {\bibinfo {volume} {75}},\
  \bibinfo {pages} {195121} (\bibinfo {year} {2007})}\BibitemShut {NoStop}%
\bibitem [{\citenamefont {Roychoudhury}\ and\ \citenamefont
  {Sanvito}(2017)}]{bib:roychoudhury17}%
  \BibitemOpen
  \bibfield  {author} {\bibinfo {author} {\bibfnamefont {S.}~\bibnamefont
  {Roychoudhury}}\ and\ \bibinfo {author} {\bibfnamefont {S.}~\bibnamefont
  {Sanvito}},\ }\href {https://link.aps.org/doi/10.1103/PhysRevB.95.085126}
  {\bibfield  {journal} {\bibinfo  {journal} {Phys. Rev. B}\ }\textbf {\bibinfo
  {volume} {95}},\ \bibinfo {pages} {085126} (\bibinfo {year}
  {2017})}\BibitemShut {NoStop}%
\bibitem [{\citenamefont {Errandonea}(1997)}]{bib:errandonea97}%
  \BibitemOpen
  \bibfield  {author} {\bibinfo {author} {\bibfnamefont {D.}~\bibnamefont
  {Errandonea}},\ }\href
  {http://www.sciencedirect.com/science/article/pii/S0375960197004192}
  {\bibfield  {journal} {\bibinfo  {journal} {Physics Letters A}\ }\textbf
  {\bibinfo {volume} {233}},\ \bibinfo {pages} {139 } (\bibinfo {year}
  {1997})}\BibitemShut {NoStop}%
\bibitem [{\citenamefont {Galanakis}\ \emph {et~al.}(2000)\citenamefont
  {Galanakis}, \citenamefont {Alouani},\ and\ \citenamefont
  {Dreyss\'e}}]{bib:galanakis00}%
  \BibitemOpen
  \bibfield  {author} {\bibinfo {author} {\bibfnamefont {I.}~\bibnamefont
  {Galanakis}}, \bibinfo {author} {\bibfnamefont {M.}~\bibnamefont {Alouani}},
  \ and\ \bibinfo {author} {\bibfnamefont {H.}~\bibnamefont {Dreyss\'e}},\
  }\href {https://link.aps.org/doi/10.1103/PhysRevB.62.6475} {\bibfield
  {journal} {\bibinfo  {journal} {Phys. Rev. B}\ }\textbf {\bibinfo {volume}
  {62}},\ \bibinfo {pages} {6475} (\bibinfo {year} {2000})}\BibitemShut
  {NoStop}%
\bibitem [{\citenamefont {Ayaz~Khan}\ \emph {et~al.}(2016)\citenamefont
  {Ayaz~Khan}, \citenamefont {Blaha}, \citenamefont {Ebert}, \citenamefont
  {Min\'ar},\ and\ \citenamefont {\ifmmode~\check{S}\else
  \v{S}\fi{}ipr}}]{bib:khan16}%
  \BibitemOpen
  \bibfield  {author} {\bibinfo {author} {\bibfnamefont {S.}~\bibnamefont
  {Ayaz~Khan}}, \bibinfo {author} {\bibfnamefont {P.}~\bibnamefont {Blaha}},
  \bibinfo {author} {\bibfnamefont {H.}~\bibnamefont {Ebert}}, \bibinfo
  {author} {\bibfnamefont {J.}~\bibnamefont {Min\'ar}}, \ and\ \bibinfo
  {author} {\bibfnamefont {O.~c.~v.}\ \bibnamefont {\ifmmode~\check{S}\else
  \v{S}\fi{}ipr}},\ }\href
  {https://link.aps.org/doi/10.1103/PhysRevB.94.144436} {\bibfield  {journal}
  {\bibinfo  {journal} {Phys. Rev. B}\ }\textbf {\bibinfo {volume} {94}},\
  \bibinfo {pages} {144436} (\bibinfo {year} {2016})}\BibitemShut {NoStop}%
\bibitem [{\citenamefont {Soler}\ \emph {et~al.}(2002)\citenamefont {Soler},
  \citenamefont {Artacho}, \citenamefont {Gale}, \citenamefont {Garc\'{\i}a},
  \citenamefont {Junquera}, \citenamefont {Ordej\'on},\ and\ \citenamefont
  {S\'anchez-Portal}}]{bib:soler02}%
  \BibitemOpen
  \bibfield  {author} {\bibinfo {author} {\bibfnamefont {J.~M.}\ \bibnamefont
  {Soler}}, \bibinfo {author} {\bibfnamefont {E.}~\bibnamefont {Artacho}},
  \bibinfo {author} {\bibfnamefont {J.~D.}\ \bibnamefont {Gale}}, \bibinfo
  {author} {\bibfnamefont {A.}~\bibnamefont {Garc\'{\i}a}}, \bibinfo {author}
  {\bibfnamefont {J.}~\bibnamefont {Junquera}}, \bibinfo {author}
  {\bibfnamefont {P.}~\bibnamefont {Ordej\'on}}, \ and\ \bibinfo {author}
  {\bibfnamefont {D.}~\bibnamefont {S\'anchez-Portal}},\ }\href
  {http://stacks.iop.org/0953-8984/14/i=11/a=302} {\bibfield  {journal}
  {\bibinfo  {journal} {Journal of Physics: Condensed Matter}\ }\textbf
  {\bibinfo {volume} {14}},\ \bibinfo {pages} {2745} (\bibinfo {year}
  {2002})}\BibitemShut {NoStop}%
\bibitem [{\citenamefont {Cuadrado}\ and\ \citenamefont
  {Cerd{\'a}}(2012)}]{bib:cuadrado12}%
  \BibitemOpen
  \bibfield  {author} {\bibinfo {author} {\bibfnamefont {R.}~\bibnamefont
  {Cuadrado}}\ and\ \bibinfo {author} {\bibfnamefont {J.~I.}\ \bibnamefont
  {Cerd{\'a}}},\ }\href {http://stacks.iop.org/0953-8984/24/i=8/a=086005}
  {\bibfield  {journal} {\bibinfo  {journal} {Journal of Physics: Condensed
  Matter}\ }\textbf {\bibinfo {volume} {24}},\ \bibinfo {pages} {086005}
  (\bibinfo {year} {2012})}\BibitemShut {NoStop}%
\bibitem [{\citenamefont {Kresse}\ and\ \citenamefont
  {Hafner}(1993)}]{bib:kresse93}%
  \BibitemOpen
  \bibfield  {author} {\bibinfo {author} {\bibfnamefont {G.}~\bibnamefont
  {Kresse}}\ and\ \bibinfo {author} {\bibfnamefont {J.}~\bibnamefont
  {Hafner}},\ }\href {https://link.aps.org/doi/10.1103/PhysRevB.47.558}
  {\bibfield  {journal} {\bibinfo  {journal} {Phys. Rev. B}\ }\textbf {\bibinfo
  {volume} {47}},\ \bibinfo {pages} {558} (\bibinfo {year} {1993})}\BibitemShut
  {NoStop}%
\bibitem [{\citenamefont {Kresse}\ and\ \citenamefont
  {Furthm\"uller}(1996)}]{bib:kresse96}%
  \BibitemOpen
  \bibfield  {author} {\bibinfo {author} {\bibfnamefont {G.}~\bibnamefont
  {Kresse}}\ and\ \bibinfo {author} {\bibfnamefont {J.}~\bibnamefont
  {Furthm\"uller}},\ }\href {\doibase 10.1103/PhysRevB.54.11169} {\bibfield
  {journal} {\bibinfo  {journal} {Phys. Rev. B}\ }\textbf {\bibinfo {volume}
  {54}},\ \bibinfo {pages} {11169} (\bibinfo {year} {1996})}\BibitemShut
  {NoStop}%
\bibitem [{\citenamefont {Bl{\"o}chl}(1994)}]{bib:bloechl94}%
  \BibitemOpen
  \bibfield  {author} {\bibinfo {author} {\bibfnamefont {P.~E.}\ \bibnamefont
  {Bl{\"o}chl}},\ }\href {\doibase 10.1103/PhysRevB.50.17953} {\bibfield
  {journal} {\bibinfo  {journal} {Phys. Rev. B}\ }\textbf {\bibinfo {volume}
  {50}},\ \bibinfo {pages} {17953} (\bibinfo {year} {1994})}\BibitemShut
  {NoStop}%
\bibitem [{\citenamefont {Steiner}\ \emph {et~al.}(2016)\citenamefont
  {Steiner}, \citenamefont {Khmelevskyi}, \citenamefont {Marsmann},\ and\
  \citenamefont {Kresse}}]{bib:steiner16}%
  \BibitemOpen
  \bibfield  {author} {\bibinfo {author} {\bibfnamefont {S.}~\bibnamefont
  {Steiner}}, \bibinfo {author} {\bibfnamefont {S.}~\bibnamefont
  {Khmelevskyi}}, \bibinfo {author} {\bibfnamefont {M.}~\bibnamefont
  {Marsmann}}, \ and\ \bibinfo {author} {\bibfnamefont {G.}~\bibnamefont
  {Kresse}},\ }\href {\doibase 10.1103/PhysRevB.93.224425} {\bibfield
  {journal} {\bibinfo  {journal} {Phys. Rev. B}\ }\textbf {\bibinfo {volume}
  {93}},\ \bibinfo {pages} {224425} (\bibinfo {year} {2016})}\BibitemShut
  {NoStop}%
\bibitem [{\citenamefont {Perdew}\ \emph {et~al.}(1997)\citenamefont {Perdew},
  \citenamefont {Burke},\ and\ \citenamefont {Ernzerhof}}]{bib:pbe96}%
  \BibitemOpen
  \bibfield  {author} {\bibinfo {author} {\bibfnamefont {J.~P.}\ \bibnamefont
  {Perdew}}, \bibinfo {author} {\bibfnamefont {K.}~\bibnamefont {Burke}}, \
  and\ \bibinfo {author} {\bibfnamefont {M.}~\bibnamefont {Ernzerhof}},\ }\href
  {https://link.aps.org/doi/10.1103/PhysRevLett.78.1396} {\bibfield  {journal}
  {\bibinfo  {journal} {Phys. Rev. Lett.}\ }\textbf {\bibinfo {volume} {78}},\
  \bibinfo {pages} {1396} (\bibinfo {year} {1997})}\BibitemShut {NoStop}%
\bibitem [{\citenamefont {Monkhorst}\ and\ \citenamefont
  {Pack}(1976)}]{bib:monk76}%
  \BibitemOpen
  \bibfield  {author} {\bibinfo {author} {\bibfnamefont {H.~J.}\ \bibnamefont
  {Monkhorst}}\ and\ \bibinfo {author} {\bibfnamefont {J.~D.}\ \bibnamefont
  {Pack}},\ }\href {\doibase 10.1103/PhysRevB.13.5188} {\bibfield  {journal}
  {\bibinfo  {journal} {Phys. Rev. B}\ }\textbf {\bibinfo {volume} {13}},\
  \bibinfo {pages} {5188} (\bibinfo {year} {1976})}\BibitemShut {NoStop}%
\bibitem [{\citenamefont {Bl\"ochl}\ \emph {et~al.}(1994)\citenamefont
  {Bl\"ochl}, \citenamefont {Jepsen},\ and\ \citenamefont
  {Andersen}}]{bib:bloechl94b}%
  \BibitemOpen
  \bibfield  {author} {\bibinfo {author} {\bibfnamefont {P.~E.}\ \bibnamefont
  {Bl\"ochl}}, \bibinfo {author} {\bibfnamefont {O.}~\bibnamefont {Jepsen}}, \
  and\ \bibinfo {author} {\bibfnamefont {O.~K.}\ \bibnamefont {Andersen}},\
  }\href {https://link.aps.org/doi/10.1103/PhysRevB.49.16223} {\bibfield
  {journal} {\bibinfo  {journal} {Phys. Rev. B}\ }\textbf {\bibinfo {volume}
  {49}},\ \bibinfo {pages} {16223} (\bibinfo {year} {1994})}\BibitemShut
  {NoStop}%
\bibitem [{\citenamefont {Fern\'andez-Seivane}\ \emph
  {et~al.}(2006)\citenamefont {Fern\'andez-Seivane}, \citenamefont {Oliveira},
  \citenamefont {Sanvito},\ and\ \citenamefont {Ferrer}}]{bib:fernandez06}%
  \BibitemOpen
  \bibfield  {author} {\bibinfo {author} {\bibfnamefont {L.}~\bibnamefont
  {Fern\'andez-Seivane}}, \bibinfo {author} {\bibfnamefont {M.~A.}\
  \bibnamefont {Oliveira}}, \bibinfo {author} {\bibfnamefont {S.}~\bibnamefont
  {Sanvito}}, \ and\ \bibinfo {author} {\bibfnamefont {J.}~\bibnamefont
  {Ferrer}},\ }\href {https://doi.org/10.1088/0953-8984/18/34/012} {\bibfield
  {journal} {\bibinfo  {journal} {Journal of Physics: Condensed Matter}\
  }\textbf {\bibinfo {volume} {18}},\ \bibinfo {pages} {7999} (\bibinfo {year}
  {2006})}\BibitemShut {NoStop}%
\bibitem [{\citenamefont {Methfessel}\ and\ \citenamefont
  {Paxton}(1989)}]{bib:methfessel89}%
  \BibitemOpen
  \bibfield  {author} {\bibinfo {author} {\bibfnamefont {M.}~\bibnamefont
  {Methfessel}}\ and\ \bibinfo {author} {\bibfnamefont {A.~T.}\ \bibnamefont
  {Paxton}},\ }\href {https://link.aps.org/doi/10.1103/PhysRevB.40.3616}
  {\bibfield  {journal} {\bibinfo  {journal} {Phys. Rev. B}\ }\textbf {\bibinfo
  {volume} {40}},\ \bibinfo {pages} {3616} (\bibinfo {year}
  {1989})}\BibitemShut {NoStop}%
\bibitem [{\citenamefont {Mostofi}\ \emph {et~al.}(2014)\citenamefont
  {Mostofi}, \citenamefont {Yates}, \citenamefont {Pizzi}, \citenamefont {Lee},
  \citenamefont {Souza}, \citenamefont {Vanderbilt},\ and\ \citenamefont
  {Marzari}}]{bib:mostofi14}%
  \BibitemOpen
  \bibfield  {author} {\bibinfo {author} {\bibfnamefont {A.~A.}\ \bibnamefont
  {Mostofi}}, \bibinfo {author} {\bibfnamefont {J.~R.}\ \bibnamefont {Yates}},
  \bibinfo {author} {\bibfnamefont {G.}~\bibnamefont {Pizzi}}, \bibinfo
  {author} {\bibfnamefont {Y.-S.}\ \bibnamefont {Lee}}, \bibinfo {author}
  {\bibfnamefont {I.}~\bibnamefont {Souza}}, \bibinfo {author} {\bibfnamefont
  {D.}~\bibnamefont {Vanderbilt}}, \ and\ \bibinfo {author} {\bibfnamefont
  {N.}~\bibnamefont {Marzari}},\ }\href
  {http://www.sciencedirect.com/science/article/pii/S001046551400157X}
  {\bibfield  {journal} {\bibinfo  {journal} {Computer Physics Communications}\
  }\textbf {\bibinfo {volume} {185}},\ \bibinfo {pages} {2309 } (\bibinfo
  {year} {2014})}\BibitemShut {NoStop}%
\bibitem [{\citenamefont {Burkert}\ \emph {et~al.}(2005)\citenamefont
  {Burkert}, \citenamefont {Eriksson}, \citenamefont {Simak}, \citenamefont
  {Ruban}, \citenamefont {Sanyal}, \citenamefont {Nordstr\"om},\ and\
  \citenamefont {Wills}}]{bib:burkert05}%
  \BibitemOpen
  \bibfield  {author} {\bibinfo {author} {\bibfnamefont {T.}~\bibnamefont
  {Burkert}}, \bibinfo {author} {\bibfnamefont {O.}~\bibnamefont {Eriksson}},
  \bibinfo {author} {\bibfnamefont {S.~I.}\ \bibnamefont {Simak}}, \bibinfo
  {author} {\bibfnamefont {A.~V.}\ \bibnamefont {Ruban}}, \bibinfo {author}
  {\bibfnamefont {B.}~\bibnamefont {Sanyal}}, \bibinfo {author} {\bibfnamefont
  {L.}~\bibnamefont {Nordstr\"om}}, \ and\ \bibinfo {author} {\bibfnamefont
  {J.~M.}\ \bibnamefont {Wills}},\ }\href
  {https://link.aps.org/doi/10.1103/PhysRevB.71.134411} {\bibfield  {journal}
  {\bibinfo  {journal} {Phys. Rev. B}\ }\textbf {\bibinfo {volume} {71}},\
  \bibinfo {pages} {134411} (\bibinfo {year} {2005})}\BibitemShut {NoStop}%
\bibitem [{\citenamefont {Gambardella}\ \emph {et~al.}(2003)\citenamefont
  {Gambardella}, \citenamefont {Rusponi}, \citenamefont {Veronese},
  \citenamefont {Dhesi}, \citenamefont {Grazioli}, \citenamefont {Dallmeyer},
  \citenamefont {Cabria}, \citenamefont {Zeller}, \citenamefont {Dederichs},
  \citenamefont {Kern}, \citenamefont {Carbone},\ and\ \citenamefont
  {Brune}}]{bib:gambardella03}%
  \BibitemOpen
  \bibfield  {author} {\bibinfo {author} {\bibfnamefont {P.}~\bibnamefont
  {Gambardella}}, \bibinfo {author} {\bibfnamefont {S.}~\bibnamefont
  {Rusponi}}, \bibinfo {author} {\bibfnamefont {M.}~\bibnamefont {Veronese}},
  \bibinfo {author} {\bibfnamefont {S.~S.}\ \bibnamefont {Dhesi}}, \bibinfo
  {author} {\bibfnamefont {C.}~\bibnamefont {Grazioli}}, \bibinfo {author}
  {\bibfnamefont {A.}~\bibnamefont {Dallmeyer}}, \bibinfo {author}
  {\bibfnamefont {I.}~\bibnamefont {Cabria}}, \bibinfo {author} {\bibfnamefont
  {R.}~\bibnamefont {Zeller}}, \bibinfo {author} {\bibfnamefont {P.~H.}\
  \bibnamefont {Dederichs}}, \bibinfo {author} {\bibfnamefont {K.}~\bibnamefont
  {Kern}}, \bibinfo {author} {\bibfnamefont {C.}~\bibnamefont {Carbone}}, \
  and\ \bibinfo {author} {\bibfnamefont {H.}~\bibnamefont {Brune}},\ }\href
  {http://science.sciencemag.org/content/300/5622/1130} {\bibfield  {journal}
  {\bibinfo  {journal} {Science}\ }\textbf {\bibinfo {volume} {300}},\ \bibinfo
  {pages} {1130} (\bibinfo {year} {2003})}\BibitemShut {NoStop}%
\bibitem [{\citenamefont {Kota}\ and\ \citenamefont
  {Sakuma}(2014)}]{bib:kota14}%
  \BibitemOpen
  \bibfield  {author} {\bibinfo {author} {\bibfnamefont {Y.}~\bibnamefont
  {Kota}}\ and\ \bibinfo {author} {\bibfnamefont {A.}~\bibnamefont {Sakuma}},\
  }\href {https://doi.org/10.7566/JPSJ.83.034715} {\bibfield  {journal}
  {\bibinfo  {journal} {Journal of the Physical Society of Japan}\ }\textbf
  {\bibinfo {volume} {83}},\ \bibinfo {pages} {034715} (\bibinfo {year}
  {2014})}\BibitemShut {NoStop}%
\bibitem [{Note1()}]{Note1}%
  \BibitemOpen
  \bibinfo {note} {In the FeAu alloy, the Au-$d$ states are mostly confined in
  a band between $-7$ and $-4$\protect \tmspace +\thinmuskip {.1667em}eV below
  the Fermi energy (see Supplementary Information Fig.~5). On the one hand,
  those states are subject to strong couplings by SOI, since $\xi _\protect
  \mathrm {Au}=615$\protect \tmspace +\thinmuskip {.1667em}meV. On the other
  hand, because of the $(\epsilon _{kn\sigma }-\epsilon _{kn'\sigma '})^{-1}$
  factors, those states have a weaker effect on the MAE for $N_e$ values close
  to charge neutrality ($N_e=19$) than for smaller $N_e$ values. E.g. a band
  filling $N_e=10$ corresponds to a downward shift of the Fermi level of
  $-3.7$\protect \tmspace +\thinmuskip {.1667em}eV, close to the Au-$d$ states.
  Therefore, fast sharp oscillations are observed in the MAE curve at
  $N_e=5-10$, while smooth behavior and apparent agreement with NSCF exists at
  $N_e>12$. Importantly, this does not mean that the Au-$d$ states have a
  negligible contribution, as evidenced by the absence of a plateau in the
  occupied states curve of the FeAu panel of Fig.~\ref {fig:fermiwinsiesta},
  which represents the $N_e=19$ case. For the FePt alloy, since the Pt-$d$ band
  is less localized in energies, the disagreement between 2PT and NSCF is
  visible throughout the $\protect \mathrm {MAE}(N_e)$ curve.}\BibitemShut
  {Stop}%
\bibitem [{\citenamefont {Pryce}(1950)}]{bib:pryce50}%
  \BibitemOpen
  \bibfield  {author} {\bibinfo {author} {\bibfnamefont {M.~H.~L.}\
  \bibnamefont {Pryce}},\ }\href {http://stacks.iop.org/0370-1298/63/i=1/a=304}
  {\bibfield  {journal} {\bibinfo  {journal} {Proceedings of the Physical
  Society. Section A}\ }\textbf {\bibinfo {volume} {63}},\ \bibinfo {pages}
  {25} (\bibinfo {year} {1950})}\BibitemShut {NoStop}%
\bibitem [{\citenamefont {Dadi}\ \emph {et~al.}(2008)\citenamefont {Dadi},
  \citenamefont {Hongjun},\ and\ \citenamefont {Myung-Hwan}}]{bib:dai08}%
  \BibitemOpen
  \bibfield  {author} {\bibinfo {author} {\bibfnamefont {D.}~\bibnamefont
  {Dadi}}, \bibinfo {author} {\bibfnamefont {X.}~\bibnamefont {Hongjun}}, \
  and\ \bibinfo {author} {\bibfnamefont {W.}~\bibnamefont {Myung-Hwan}},\
  }\href {https://onlinelibrary.wiley.com/doi/abs/10.1002/jcc.21011} {\bibfield
   {journal} {\bibinfo  {journal} {Journal of Computational Chemistry}\
  }\textbf {\bibinfo {volume} {29}},\ \bibinfo {pages} {2187} (\bibinfo {year}
  {2008})}\BibitemShut {NoStop}%
\end{thebibliography}%

\end{document}